\documentclass[hidelinks,twocolumn, prl, superscriptaddress,a4paper]{revtex4}

\usepackage[colorlinks=true,allcolors=blue]{hyperref}

\usepackage{graphicx}% Include figure files
\usepackage{dcolumn}% Align table columns on decimal point
\usepackage{orcidlink}
\usepackage{csquotes}
\usepackage{float}
\usepackage{amsmath,amssymb}
\usepackage{braket}
\usepackage{bm}% bold math
\begin{document}

\preprint{APS123-QED}

%\preprint{}

\title{Generalization of Modular Spread Complexity for Non-Hermitian Density Matrices}


% \affiliation can be followed by \email, \homepage, \thanks as well..

\author{Aneek Jana\,\orcidlink{0009-0001-1097-4250}}

\affiliation{Center for High Energy Physics, Indian Institute of Science, Bangalore}

\author{Maitri Ganguli\,\orcidlink{0009-0009-4701-2459}}
%\thanks{}
\affiliation{Department of Physics, Indian Institute of Science, Bangalore}


\date{\today}

\begin{abstract}
In this work we generalize the concept of modular spread complexity to the cases where the reduced density matrix is non-Hermitian. This notion of complexity and associated Lanczos coefficients contain richer information than the pseudo-entropy, which turns out to be one of the first Lanczos coefficients. We also define the quantity pseudo-capacity which generalizes capacity of entanglement, and corresponds to the early modular-time measure of pseudo-modular complexity.
We describe how pseudo-modular complexity can be calculated using a slightly modified bi-Lanczos algorithm. Alternatively, the (complex) Lanczos coefficients can also be obtained from the analytic expression of the pseudo-R\'enyi entropy, which can then be used to calculate the pseudo-modular spread complexity. We show some analytical calculations for 2-level systems and 4-qubit models and then do numerical investigations on the quantum phase transition of transverse field Ising model, from the (pseudo) modular spread complexity perspective. As the final example, we consider the \(3d\) Chern-Simon gauge theory with Wilson loops to understand the role of topology on modular complexity. The concept of pseudo-modular complexity introduced here can be a useful tool for understanding phases and phase transitions in quantum many body systems, quantum field theories and holography.

\end{abstract}

\maketitle

\section{Introduction}\label{sec: Intro}
The study of quantum complexity in the Krylov space has become an interesting and exciting research direction in recent years. Dynamics in the Krylov space~\cite{Nandy:2024htc} has been investigated in diverse fields, starting from quantum many-body physics to quantum circuits to high energy physics~\cite{Parker_2019,Camargo:2024deu,Balasubramanian:2022tpr,Rabinovici_2021,Bhattacharjee_2023,Chapman:2024pdw,Bhattacharjee_2022,Bhattacharya:2023xjx,Bhattacharya:2022gbz,Bhattacharjee_2024,Bhattacharya:2023zqt,Bhattacharya:2023yec,Bhattacharya:2024hto,suchsland2023krylovcomplexitytrottertransitions,Sahu:2024urf,Avdoshkin_2024,Camargo_2023,Vasli_2024,he2024probingkrylovcomplexityscalar,PhysRevD.104.L081702,Kundu_2023,Bhattacharya:2024szw}. Complexity of time-evolution of both operators and states are the two things that have been studied in the above mentioned literature. In the case of operators, the time evolution is generated by Liouvillian in the Heisenberg picture, and for the case of states, the time evolution is generated by usual Hamiltonian in the Schrodinger picture. In the context of states, Krylov complexity is also known as the spread complexity, which is defined as the minimal amount of spread of the wave function in the Hilbert space and this minimization is achieved in the Krylov space in a finite amount of time~\cite{Balasubramanian:2022tpr}. In the recent works, it has been observed that the spread complexity of thermofield double state (TFD) is an important quantity that can distinguish chaotic dynamics from integrable dynamics~\cite{Camargo:2024deu}, in particular, the peak height in the TFD spread complexity can be treated as an order parameter in the chaotic to integrable phase transitions~\cite{Baggioli:2024wbz,Ganguli:2024myj}.

The idea of modular complexity has been introduced lately~\cite{caputa_modular} with the goal to define a notion of complexity capturing the entanglement structure of a general bipartite quantum state. Entanglement entropy only captures a part of the information contained in the full entanglement spectrum (that is, the eigenvalue spectrum of the reduced density matrix) and the concept of modular complexity makes the picture complete since it utilizes the entire entanglement spectrum. One starts with a density matrix and first, constructs a TFD-like purifier of that density matrix which will act as the initial state. On the other hand, this initial state has to be evolved in modular time using the modular Hamiltonian (also known as the entanglement Hamiltonian)~\cite{Dalmonte_2022}, defined as the negative logarithm of the density matrix. Then one calculates the modular Lanczos coefficients and the modular complexity is defined as the usual Krylov complexity corresponding to the modular Hamiltonian evolution. Surprisingly, in this endeavor to combine the ideas of entaglement and complexity, the first Lanczos coefficient \(a_0\) turns out to be the von-Neumann entropy \(S_E\)~\cite{headrick2019lecturesentanglemententropyfield,Rangamani_2017} and the next Lanczos coefficient \(b_1\) becomes the square root of capacity of entaglement \(C_E\)~\cite{de_Boer_2019,Wei_2023}. At this level itself one can gather non-trivial information, since the early time complexity is always proportional to \(b_1^2\), which is nothing but \(C_E\). Therefore, we can consider capacity of entanglement as the early time measure of modular complexity. This observation perfectly goes with our intuitive idea of complexity since modular complexity vanishes identically for unentangled states as well as maximally entangled states (degenerate entanglement spectrum).

On the other hand, a new generalization of entanglement entropy has been proposed recently which is known as the pseudo-entropy~\cite{Nakata_2021}, the main motivation being to generalize holographic entanglement entropy to time-dependent Euclidean \(AdS\) background. In this case, naturally, one has two differently prepared states \(\ket{\psi}\) and \(\ket{\varphi}\) and then the transition matrix, generalization of density matrix, is defined by \(\rho^{\psi|\varphi} = \ket{\psi}\bra{\varphi}/\braket{\varphi|\psi}\). Such construction is also applicable to cases where the final state \(\ket{\varphi}\) is post-selected from the initial state \(\ket{\psi}\), and the transition matrix plays the role of density matrix while computing the weak values of the operators. Then pseudo-entropy is defined as the von-Neumann entropy of the reduced transition matrix. Starting from its proposal, pseudo-entropy has been investigated in vast literature in various contexts, for example in free QFTs~\cite{Mollabashi_2021}, in entanglement phase transition in holography~\cite{Kanda_2024}, pseudo-entropy in joining local quenches~\cite{Shinmyo_2024}, in \(U(1)\) gauge theory~\cite{Mukherjee_2022}, in topological field theoris~\cite{Nishioka_2021,caputa2024musingssvdpseudoentanglement}, in the context of \(dS/CFT\) correspondence~\cite{Doi_2023} etc.

Motivated by these developments, in this work, we have introduced the notion of modular complexity applicable for the cases where the density matrix is non-Hermitian. In this context, we call this quantity pseudo-modular spread complexity, and it reduces to the modular spread complexity in the Hermitian limit. This will allow us to study the notion of complexity in a variety of physical settings in which cases the behavior pseudo-entropy has already been analysed. Just like the capacity of entanglement in the Hermitian cases, we have to define the notion of pseudo-capacity when the reduced density matrix is non-Hermitian. It turns out that the early-time pseudo-modular spread complexity is proportional to the absolute value of the pseudo-capacity. The full complexity profile of course depends on the higher cumulants of the (pseudo) entanglement spectrum. In this paper, we exemplify these ideas using simple qubit systems, transverse field Ising model (TFIM) and \(3d\) Chern-Simons gauge theory with Wilson loops.

We begin by reviewing the concept of modular spread complexity in Sec.~\ref{sec: modular}. In Sec.~\ref{sec: psedu-modular} we generalize the notion of modular spread complexity to non-Hermitian density matrices and introduce the concept of right and left pseudo-modular spread complexities. We also calculate some explicit results for 2-level systems, in this section. In Sec.~\ref{sec: qubit examples}, we analyse qubit systems and quantum phase transition in transverse field Ising model (TFIM) from the perspective of modular and pseudo-modular spread complexity. In Sec.~\ref{sec: Chern-Simons gauge theory}, we discuss the \(3d\) Chern-Simons gauge theory with Wilson loops, which is a topological field theory. Finally, we conclude with some discussions and future directions in Sec.~\ref{sec: discussion}.

\section{Modular Spread Complexity}\label{sec: modular}
Modular spread complexity is defined for usual density matrices which are Hermitian. Modular spread complexity essentially depends on the eigenvalues or 'entanglement spectrum' of the density matrix, which can be thought of as a reduced density matrix coming from a full pure density matrix of a bigger Hilbert space by tracing out certain subsystem. In any case, we can write the spectral decomposition of the density matrix as,
\begin{equation}\label{eq: density matrix}
    \rho = \sum_\alpha \lambda_\alpha \ket{\alpha}\bra{\alpha}
\end{equation}
from \(\text{Tr}\rho=1\) we have \(\sum_\alpha\lambda_\alpha = 1\) and from positivity \(\lambda_\alpha \geq 0\). Then we have the modular Hamiltonian \(H_M\) as,
\begin{equation}
    H_M = - \log \rho = \sum_\alpha \mathcal E_\alpha \ket{\alpha}\bra{\alpha}
\end{equation}
therefore, the modular energy spectrum \(\mathcal E_\alpha\) depends on the entanglement spectrum \(\lambda_\alpha\) by \(\lambda_\alpha = e^{-\mathcal E_\alpha}\). 

To compute spread complexity, one starts with the canonical purification of the density matrix in Eq.(\ref{eq: density matrix}),
\begin{equation}
    \ket{\Psi} = \sum_\alpha \sqrt{\lambda_\alpha} \ket{\alpha}^{(1)}\ket{\alpha}^{(2)} = \sum_\alpha e^{-\mathcal E_\alpha/2} \ket{\alpha}^{(1)}\ket{\alpha}^{(2)}
\end{equation}
where \(\ket{\alpha}^{(1)}\) and \(\ket{\alpha}^{(2)}\) belongs to two different copies of the same Hilbert space where \(\ket{\alpha}\) belongs. For non-trivial modular time evolution the state \(\ket{\Psi}\) is evolved with \(H_M\otimes\mathbb I\). Therefore, the modular time evolved state is,
\begin{equation}
\begin{aligned}
    \ket{\Psi(s)} &= e^{-i s H_M\otimes\mathbb I } \ket{\Psi}\\
    &= \sum_\alpha e^{-i s \mathcal E_\alpha} e^{- \mathcal E_\alpha/2} \ket{\alpha}^{(1)}\ket{\alpha}^{(2)}
\end{aligned}
\end{equation}
Modular spread complexity corresponds to Krylov complexity of the above mentioned modular time evolution. It can be observed that the state in the second copy of the Hilbert space practically remains a spectator in the whole evolution, so we can obtain same complexity if one considers the following evolution,
\begin{equation}\label{eq: modular time evolution}
\begin{aligned}
    &\ket{\Psi} = \sum_\alpha e^{-\mathcal E_\alpha/2} \ket{\alpha}\\
    &\ket{\Psi(s)} = e^{-i s H_M} \ket{\Psi} = \sum_\alpha e^{-i s \mathcal E_\alpha} e^{- \mathcal E_\alpha/2} \ket{\alpha}
\end{aligned}
\end{equation}

First, orthonormal Krylov basis vectors \(\{\ket{K_n}:n\geq0\}\) are created using the Lanczos algorithm,
\begin{itemize}
    \item Define \(\ket{K_0} = \ket{\Psi}\) and \(a_0 = \bra{K_0} H_M \ket{K_0}\).
    \item For \(n=1\),
    \begin{equation}
    \begin{aligned}
        &\ket{A_1} = H_M \ket{K_0} - a_0 \ket{K_0} \\
        &b_1^2 = \braket{A_1|A_1}\\
        &\ket{K_1} = \ket{A_1}/b_1\\
        & a_1 = \bra{K_1} H_M \ket{K_1}
    \end{aligned}
    \end{equation}
    \item For \(n>1\),
    \begin{equation}
    \begin{aligned}
        &\ket{A_n} = (H_M - a_{n-1}) \ket{K_{n-1}} - b_{n-1} \ket{K_{n-2}}\\
        &b_n^2 = \braket{A_n|A_n}\\
        &\ket{K_n} = \ket{A_n}/b_n\\
        & a_n = \bra{K_n} H_M \ket{K_n}
    \end{aligned}
    \end{equation}
\end{itemize}

By direct calculation one obtains the following imporatant insights into the modular Lanczos coefficients,
\begin{equation}\label{eq: SE and CE from a0 and b1^2}
    \begin{aligned}
        &a_0 = \sum_\alpha \mathcal E_\alpha e^{-\mathcal E_\alpha} = - \sum_\alpha \lambda_\alpha \log\lambda_\alpha = S_E\\
        & b_1^2 = \sum_\alpha \mathcal E_\alpha^2 e^{-\mathcal E_\alpha} - (\sum_\alpha \mathcal E_\alpha e^{-\mathcal E_\alpha})^2\\
        &=  \sum_\alpha \lambda_\alpha \log^2\lambda_\alpha - (\sum_\alpha \lambda_\alpha \log\lambda_\alpha)^2 = C_E   
    \end{aligned}
\end{equation}
where \(S_E\) is the entanglement entropy and \(C_E\) is the capacity of entanglement.
 
The modular Hamiltonian is cast into a tridiagonal form in the Krylov basis since,
\begin{equation}\label{eq: tridiag H_M}
    H_M \ket{K_n} = a_n \ket{K_n} + b_n \ket{K_{n-1}} + b_{n+1} \ket{K_{n+1}}
\end{equation}

Next, the evolved state \(\ket{\Psi(s)}\) is expanded in the Krylov basis,
\begin{equation}
    \ket{\Psi(s)} = \sum_n \phi_n(s) \ket{K_n}
\end{equation}

From the Schrodinger equation \(i\partial_s \ket{\Psi(s)} = H_M \ket{\Psi(s)}\) and Eq.(\ref{eq: tridiag H_M}), the following recursive differential equation can be derived for \(\phi_n(s)\),
\begin{equation}\label{eq: recursive diff eq}
    i\partial_s \phi_n(s) = a_n \phi_n (s) + b_{n+1} \phi_{n+1}(s) + b_n \phi_{n-1}(s)
\end{equation}
with the initial condition \(\phi_n(0) = \delta_{n,0}\).

Modular spread complexity is defined as the average position in the Krylov basis,
\begin{equation}\label{eq: complexity defn}
    \mathcal C(s) = \sum_n n |\phi_n(s)|^2
\end{equation}

It can be shown that the early-time behavior of complexiity is quadratic,
\begin{equation}
    \mathcal C(s) \sim b_1^2 s^2 = C_E s^2
\end{equation}
therefore, capacity of entanglement \(C_E\) can be understood as the early time measure of modular complexity. On the other hand, from Eq.(\ref{eq: modular time evolution}) we can see that the time evolution becomes trivial if the spectrum is fully degenerate/maximally entangled case or singleton/unentagled case and hence modular complexity \(\mathcal C(s)\) vanishes identically (no non-trivial dynamics in the Krylov space). It is not hard to see that \(C_E\) also vanishes precisely in these two cases.

From Eq.(\ref{eq: recursive diff eq}) and Eq.(\ref{eq: complexity defn}), it can be understood that the knowledge of Lanczos coefficients \(a_n\) and \(b_n\) is sufficient for the evaluation of complexity. Keeping this in mind, we mentions an alternative method of obtaining the Lanczos coefficients that does not use the Lanczos algorithm but yields the Lanczos coefficients using the moments of the survival amplitude,
\begin{equation}
    S(s) = \braket{\Psi(s)|\Psi(0)}
\end{equation}

The moments are defined as,
\begin{equation}\label{eq: moment defn}
    m^{(k)} = \frac{1}{i^k}\frac{d^k}{ds^k}S(s)|_{s=0}
\end{equation}

The relation between moments and first few Lanczos coefficients are,
\begin{equation}
\begin{aligned}\label{eq: lanc from mom}
    &a_0 = m^{(1)}\\
    &b_1^2 = m^{(2)} - a_0^2\\
    &a_1= \frac{m^{(3)} - a_0^3 - 2 a_0 b_1^2}{b_1^2}
\end{aligned}
\end{equation}

In the present case,
\begin{equation}\label{eq: modular survival}
    S(s) = \sum_\alpha e^{-\mathcal E_\alpha(1-i s)} = \sum_\alpha \lambda_\alpha^{1-is} = \text{Tr}\rho^{1-is}
\end{equation}
which can be obtained from the analytic continuation of \(n\)th R\'enyi entropy. Using Eq.(\ref{eq: moment defn})-(\ref{eq: modular survival}) we can recover Eq.(\ref{eq: SE and CE from a0 and b1^2}). Suppose, we define,
\begin{equation}
    Z(n) = \text{Tr}\rho^n
\end{equation}
Therefore, \(S(s) = Z(1-is)\) and from Eq.(\ref{eq: moment defn}), we can find,
\begin{equation}
    m^{(k)} = (-1)^k \frac{\partial^k}{\partial n^k}Z(n)|_{n=1}
\end{equation}
All the Lanczos coefficients can be determined from the knowledge of the moments~\cite{Nandy:2024htc}, which suffice to calculate the complexity profile as a function of modular time.

\section{Pseudo-Modular Spread Complexity}\label{sec: psedu-modular}

In this section we generalize the notion of modular complexity for the non-Hermitian density matrix cases. Non-Hermitian reduced density matrix can be obtained by taking partial trace on a generalized density matrix, known as the transition matrix which involves two non-orthogonal states \(\ket{\psi}\) and \(\ket{\varphi}\),
\begin{equation}
    \rho^{\psi|\varphi} = \frac{\ket{\psi}\bra{\varphi}}{\braket{\varphi|\psi}}
\end{equation}
Reduced density matrix for subsystem \(A\) can be obtained by tracing out its complement \(B=A^c\),
\begin{equation}
    \rho^{\psi|\varphi}_A = \text{Tr}_B (\rho^{\psi|\varphi})
\end{equation}
In general, this operator is not Hermitian and has complex eigenvalues. The corresponding von Neumann entropy is known as the \textit{pseudo-entropy},
\begin{equation}
    S_E^{\psi|\varphi} = - \text{Tr}_A(\rho^{\psi|\varphi}_A\log \rho^{\psi|\varphi}_A)
\end{equation}
Analogous to the Hermitian case, we define the \textit{pseudo-capacity},
\begin{equation}
    C_E^{\psi|\varphi} = \text{Tr}_A(\rho^{\psi|\varphi}_A\log^2 \rho^{\psi|\varphi}_A) - (\text{Tr}_A(\rho^{\psi|\varphi}_A\log \rho^{\psi|\varphi}_A))^2
\end{equation}

Nexr we introduce the general formalism for obtaining modular complexity for non-Hermitian density matrices.

\subsection{A. General Formalism}

Let's now consider the density matrix \(\rho\) to be non-Hermitian \(\rho^\dagger \neq \rho\). We can write its spectral decomposition as,
\begin{equation}
    \rho = \sum_\alpha \lambda_\alpha \ket{\alpha}_R \bra{\alpha}_L
\end{equation}
where \(\lambda_\alpha\) now can be complex numbers still satisfying \(\sum_\alpha \lambda_\alpha = 1\), \(\ket{\alpha}_R\) and \(\bra{\alpha}_L\) are the right and left eigenvectors of \(\rho\) respectively, which are bi-orthonormal, that is \({}_L\braket{\alpha|\alpha'}_R = \delta_{\alpha,\alpha'}\).
\begin{equation}
\begin{aligned}
    &\rho \ket{\alpha}_R = \lambda_\alpha \ket{\alpha}_R\\
    &\bra{\alpha}_L \rho = \lambda_\alpha \bra{\alpha}_L
\end{aligned}
\end{equation}
\(\ket{\alpha}_L\) can be understood as the right eigenvector of \(\rho^\dagger\) with eigenvalue \(\lambda_\alpha^*\),
\begin{equation}
    \rho^\dagger \ket{\alpha}_L = \lambda_\alpha^* \ket{\alpha}_L
\end{equation}
We have the pseudo-modular Hamiltonian as \(H_M = - \log \rho\) which is now non-Hermitian, \(H_M^\dagger \neq H_M\). \(H_M\) has eigenvalues \(\mathcal E_\alpha = - \log\lambda_\alpha\) and eigenvectors \(\ket{\alpha}_R\). \(H_M^\dagger\) has eigenvalues \(\mathcal E_\alpha^*\) and eigenvectors \(\ket{\alpha}_L\).

We will be interested in the complexity of modular evolution with both \(H_M\) and \(H_M^\dagger\). First we define two initial states,
\begin{equation}
    \begin{aligned}
        &\ket{\psi_R} = \sum_\alpha e^{-\mathcal E_\alpha/2} \ket{\alpha}_R\\
         &\ket{\psi_L} = \sum_\alpha e^{-\mathcal E_\alpha^*/2} \ket{\alpha}_L
    \end{aligned}
\end{equation}
clearly, \({}_L\braket{\alpha|\alpha}_R = 1\).

Modular time-evolved states are,
\begin{equation}
    \begin{aligned}
        &\ket{\psi_R(s)} = e^{-i H_M s}\ket{\psi_R}=\sum_\alpha e^{-i\mathcal E_\alpha s} e^{-\mathcal E_\alpha/2} \ket{\alpha}_R\\
         &\ket{\psi_L(s)} = e^{-i H_M^\dagger s}\ket{\psi_L}=\sum_\alpha e^{-i\mathcal E_\alpha^* s} e^{-\mathcal E_\alpha^*/2} \ket{\alpha}_L
    \end{aligned}
\end{equation}
The idea is to obtain the complexities of the above two evolutions simultaneously by constructing a two bi-orthogonal sets of Krylov basis vectors. This can be achieved easily using the bi-Lanczos algorithm, which we now describe.
\begin{itemize}
    \item Define \(\ket{P_0}=\ket{\psi_R}\), \(\ket{Q_0}=\ket{\psi_L}\) and \(a_0 = \bra{Q_0}H_M\ket{P_0}\).
    \item For \(n=1\),
    \begin{equation}
    \begin{aligned}
        &\ket{A_1} = H_M \ket{P_0} - a_0 \ket{P_0}\\
        &\ket{B_1} = H_M^\dagger \ket{Q_0} - a_0^* \ket{Q_0}\\
        & b_1^2 = \braket{B_1|A_1}\\
        &\ket{P_1} = \ket{A_1}/b_1,\,\,\,\ket{Q_1} = \ket{B_1}/b_1^*\\
        &a_1 = \bra{Q_1}H_M \ket{P_1}
    \end{aligned}
    \end{equation}
    \item For \(n>1\),
    \begin{equation}
    \begin{aligned}
        &\ket{A_n} = (H_M - a_{n-1}) \ket{P_{n-1}} - b_{n-1} \ket{P_{n-2}}\\
        &\ket{B_n} = (H_M^\dagger - a_{n-1}^*) \ket{Q_{n-1}} - b_{n-1}^* \ket{Q_{n-2}}\\
        & b_n^2 = \braket{B_n|A_n}\\
        &\ket{P_n} = \ket{A_n}/b_n,\,\,\,\ket{Q_n} = \ket{B_n}/b_n^*\\
        &a_n = \bra{Q_n}H_M \ket{P_n}
    \end{aligned}
    \end{equation}
\end{itemize}
By construction, \(\braket{Q_n|P_m} = \delta_{nm}\) and the Lanczos coefficients \(a_n,b_n\) can be complex numbers. Analgous to the Hermitian case, \(a_0\) turns out to be the pseudo-entropy and \(b_1^2\) turns out to be the pseudo-capacity of the non-Hermitian density matrix \(\rho\).

To compute complexity, the states \(\ket{\psi_R (s)}\) and \(\ket{\psi_L (s)}\) are expanded in the bases \(\{\ket{P_n}\}\) and \(\{\ket{Q_n}\}\) respectively,
\begin{equation}
    \begin{aligned}
        &\ket{\psi_R(s)} = \sum_n \phi^R_n (s)\ket{P_n};\,\,\,\phi_n^R(s) = \braket{Q_n|\psi_R(s)}\\
        &\ket{\psi_L(s)} = \sum_n \phi^L_n (s)\ket{Q_n};\,\,\,\phi_n^L(s) = \braket{P_n|\psi_L(s)}
    \end{aligned}
\end{equation}
This allows us to define \textit{right modular spread complexity} and \textit{left modular spread complexity} as,
\begin{equation}
    \begin{aligned}
        &\mathcal C_R(s) = \frac{\sum_n n |\phi_n^R(s)|^2}{\sum_n |\phi_n^R(s)|^2}\\
        &\mathcal C_L(s) = \frac{\sum_n n |\phi_n^L(s)|^2}{\sum_n |\phi_n^L(s)|^2}
    \end{aligned}
\end{equation}
The way the formalism is set up, it is not hard to see that the right modular complexity and the left modular complexity will be equal to each other when the set \(\{\mathcal E_\alpha\}\) is equal to the set \(\{\mathcal E_\alpha^*\}\). This will be true if the density matrix is real, that is \(\rho = \rho^*\) in some basis.

Recursive differential equations satisfied by \(\phi_n^R(s)\) and \(\phi_n^L(s)\) are,
\begin{equation}\label{eq: rec. diff eqn pseudo}
    \begin{aligned}
        &i\partial_s \phi_n^R (s) = a_n \phi_n^R (s) + b_{n+1}\phi_{n+1}^R (s) + b_n \phi_{n-1}^R (s)\\
        &i\partial_s \phi_n^L (s) = a_n^*\phi_n^L (s) + b_{n+1}^*\phi_{n+1}^L (s) + b_n^* \phi_{n-1}^L (s)
    \end{aligned}
\end{equation}
with \(\phi_n^R(0) = \phi_n^L(0) = \delta_{n,0}\).

The Lanczos coefficients can also be obtained from the moment recursion algorithm using the moments of the following \textit{right survival amplitude},
\begin{equation}
    S_R (s) = \braket{\psi_L(s)|\psi_R(0)} = \bra{\psi_L} e^{+i H_M s} \ket{\psi_R}
\end{equation}
\textit{Right modular form factor} is defined by,
\begin{equation}
    \text{MoFF}_R (s) = |S_R(s)|^2
\end{equation}
Similarly \textit{left survival amplitude} and \textit{left modular form factor},
\begin{equation}
    \begin{aligned}
        &S_L (s) = \braket{\psi_R(s)|\psi_L(0)}\\
        &\text{MoFF}_L (s) = |S_L(s)|^2
    \end{aligned}
\end{equation}

The right survival amplitude can be understood as,
\begin{equation}
    S_R (s) = \text{Tr}\rho^{1-is} = Z(1-is)
\end{equation}
Therefore, the (right) moments can be obtained by,
\begin{equation}
    m_R^{(k)} = \frac{1}{i^k}\frac{d^k}{ds^k}S_R(s)|_{s=0} = (-1)^k \frac{\partial^k}{\partial n^k}Z(n)|_{n=1}
\end{equation}

The Lanczos coefficients \(\{a_n\}\) and \(\{b_n\}\) can be obtained from the above moments using the moment recursion algorithm.

The left-moments are just the complex conjugates of the right-moments,
\begin{equation}
    m_L^{(k)} = (m_R^{(k)})^*
\end{equation}
and they will correspond to complex conjugated Lanczos coefficients \(\{a_n^*\}\) and \(\{b_n^*\}\), which dictate the profile of left pseudo-modular spread complexity.

\subsection{B. Explicit Results for 2-level Systems}
Now we show some explicit results for pseudo-modular spread complexity for 2-level systems. The density matrix can be written as,
\begin{equation}
    \rho = \lambda_0 \ket{0_R}\bra{0_L}+\lambda_1 \ket{1_R}\bra{1_L}
\end{equation}
with \(\lambda_0 + \lambda_1 = 1\). Such non-Hermitian density matrix can arise, for example, if we consider the transition matrix \(\rho^{\psi|\varphi} = \ket{\psi}\bra{\varphi}/\braket{\varphi|\psi}\) between two different GHZ states,
\begin{equation}
\begin{aligned}
    &\ket{\psi} = \sqrt{p_1} \ket{000} + \sqrt{1-p_1} \ket{111}\\
    &\ket{\varphi} = \sqrt{p_2} \ket{000} + \sqrt{1-p_2} \, e^{i\phi}\ket{111}
\end{aligned}
\end{equation}
and then trace-out the first two qubits,
\begin{equation}
\begin{aligned}
    \rho_A^{\psi|\varphi} &= \text{Tr}_B \rho^{\psi|\varphi}\\
    &= \frac{\sqrt{p_1 p_2}\ket{0}\bra{0}+\sqrt{(1-p_1) (1-p_2)}e^{-i\phi}\ket{1}\bra{1}}{\sqrt{p_1 p_2}+\sqrt{(1-p_1) (1-p_2)}e^{-i\phi}} 
\end{aligned}
\end{equation}

We take \(\lambda_0 = p + i q \) and \(\lambda_1 = 1 - p - i q \), where \(p,q\) are real numbers. Modular energy spectrum is, \(\mathcal E_0 = - \log \lambda_0\) and \(\mathcal E_1 = - \log \lambda_1\).

Krylov basis vectors are,
\begin{equation}
    \begin{aligned}
        &\ket{P_0} = e^{-\mathcal{E}_0/2} \ket{0_R} + e^{-\mathcal{E}_1/2 } \ket{1_R}\\
        &\ket{P_1} =  (e^{-\mathcal{E}_0/2} ( \mathcal{E}_0 - a_0) \ket{0_R} + e^{-\mathcal{E}_1/2} ( \mathcal{E}_1 - a_0) \ket{1_R})/b_1\\
        &\ket{Q_0} = e^{-\mathcal{E}_0^*/2} \ket{0_L} + e^{-\mathcal{E}_1^*/2 } \ket{1_L}\\
        &\ket{Q_1} =  (e^{-\mathcal{E}_0^*/2} ( \mathcal{E}_0^* - a_0^*) \ket{0_L} + e^{-\mathcal{E}_1^*/2} ( \mathcal{E}_1^* - a_0^*) \ket{1_L})/b_1^*
    \end{aligned}
\end{equation}
with,
\begin{equation}
    \begin{aligned}
        &a_0 = - \lambda_0 \log \lambda_0 - \lambda_1 \log \lambda_1\\
        &b_1^2 =  \lambda_0 (-\log\lambda_0 - a_0)^2 + \lambda_1 (-\log\lambda_1 - a_0)^2\\
        &a_1 = -\frac{\lambda_0\log\lambda_0 (-\log\lambda_0 - a_0)^2 + \lambda_1\log\lambda_1 (-\log\lambda_1 - a_0)^2}{b_1^2}
    \end{aligned}
\end{equation}
Time evolved states are,
\begin{equation}
    \begin{aligned}
         &\ket{\psi_R (s)}
     = e^{-\mathcal{E}_0/2} e^{-i \mathcal{E}_0 s} \ket{0_R} + e^{-\mathcal{E}_1/2} e^{-i\mathcal{E}_1 s}\ket{1_R}\\
     &\ket{\psi_L (s)}
     = e^{-\mathcal{E}_0^*/2} e^{-i \mathcal{E}_0^* s} \ket{0_L} + e^{-\mathcal{E}_1^*/2} e^{-i\mathcal{E}_1^* s}\ket{1_L}
    \end{aligned}
\end{equation}
and Krylov basis wave-functions,
\begin{equation}
    \begin{aligned}
    &\phi_0^R (s) = e^{-\mathcal{E}_0} e^{-i s \mathcal{E}_0} + e^{-\mathcal{E}_1} e^{-i s \mathcal{E}_1}\\
    &\phi_1^R (s) = \frac{1}{b_1} (e^{-\mathcal{E}_0} ( \mathcal{E}_0 - a_0) e^{-i s \mathcal{E}_0}  + e^{-\mathcal{E}_1} ( \mathcal{E}_1 - a_0) e^{-i s \mathcal{E}_1} )\\
    &\phi_0^L (s) = e^{-\mathcal{E}_0^*} e^{-i s \mathcal{E}_0^*} + e^{-\mathcal{E}_1^*} e^{-i s \mathcal{E}_1^*}\\
    &\phi_1^L (s) = \frac{1}{b_1^*} (e^{-\mathcal{E}_0^*} ( \mathcal{E}_0^* - a_0^*) e^{-i s \mathcal{E}_0^*}  + e^{-\mathcal{E}_1^*} ( \mathcal{E}_1^* - a_0^*) e^{-i s \mathcal{E}_1^*} )
    \end{aligned}
\end{equation}

Finally we can write the left and right modular complexity for this two level system as;

\begin{equation}
   \mathcal C_R (s) = \frac{| \phi_1^R (s) | ^2}{| \phi_0^R (s) | ^2 + | \phi_1^R (s) | ^2}
\end{equation}

\begin{equation}
    \mathcal C_L (s) = \frac{| \phi_1^L (s) | ^2}{| \phi_0^L (s) | ^2 + | \phi_1^L (s) | ^2}
\end{equation}

\begin{figure}
    \centering
    \includegraphics[width=0.48\linewidth]{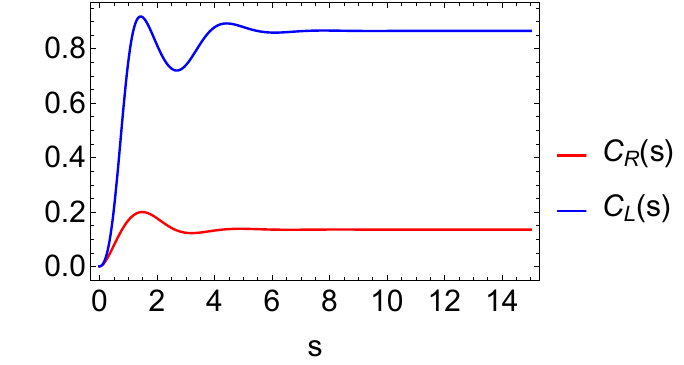}
    \includegraphics[width=0.48\linewidth]{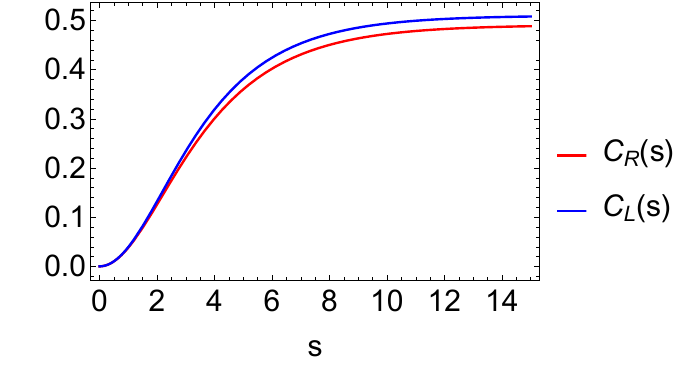}
    \includegraphics[width=0.95\linewidth]{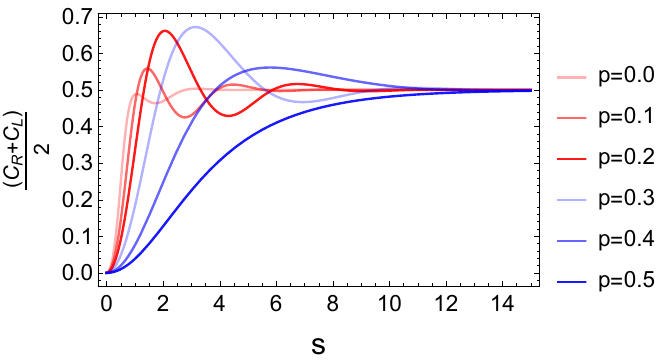}
    \caption{\textbf{2-level pseudo-modular spread complexity.} The top panel shows the right and left pseudo-modular spread complexity of 2-level system with \(p=0.1,q=0.1\) (left) and \(p=0.49,q=0.1\) (right). For \(p=0.5\), the right and left complexities coincide (not shown).
    The bottom plot shows the average of right and left modular spread complexities for fixed \(q=0.1\) and different values of \(p\). The initial growth and saturation kind of profile is absent for modular complexity of 2-level system which is always oscillating.}
    \label{fig: pseudo-modular 2-level}
\end{figure}

In Fig.~\ref{fig: pseudo-modular 2-level} shows the plots for the pseudo-modular complexity profiles for different choices of \(p\) and fixed \(q=0.1\).

\subsection{C. Level-2 Pseudo-modular spread complexity}

Here we define the notion of level-2 pseudo-modular spread complexity that can be applied to any non-Hermitian density matrix of \(\geq 2\). At first, we shall calculate the first few Lanczos coefficients \(a_0,b_1\) and \(a_1\) which are complex in-general. Then we define level-2 version of complexity by solving the following coupled differential equations,
\begin{equation}
  i \frac{d}{ds}\ket{\psi_R(s)} = i \frac{d}{ds} \begin{pmatrix}
        \phi_0^R (s) \\
        \phi_1^R (s)
    \end{pmatrix} =
    \begin{pmatrix}
        a_0 & b_1 \\
        b_1 & a_1
    \end{pmatrix}
     \begin{pmatrix}
        \phi_0^R (s) \\
        \phi_1^R (s)
    \end{pmatrix}  
\end{equation}
with initial conditions \(\phi_0^R(0)=1\) and \(\phi_1^R(0)=0\).

The general solution is,
\begin{equation}
    \ket{\psi_R(s)} = c_1 e^{- i E_1 s}\ket{\psi_1} + c_2 e^{- i E_2 s}\ket{\psi_2}
\end{equation}
where \(\ket{\psi_1}\) and \(\ket{\psi_2}\) are the eigenvectors of the \(2\times 2\) matrix with eigenvalues \(E_1\) and \(E_2\) respectively. The specific solution with given boundary conditions are,
\begin{equation}
    \ket{\psi_R(s)}
    = \frac{1}{E_2 - E_1}
    \begin{pmatrix}
        (E_2 - a_0)e^{-i E_1 s} - (E_1 - a_0)e^{-i E_2 s}\\
        -b_1 e^{-i E_1 s} + b_1 e^{-i E_2 s}
    \end{pmatrix}
\end{equation}
here,
\begin{equation}
    \begin{aligned}
        E_1 = \frac{1}{2}\left(a_0 + a_1 - \sqrt{(a_0-a_1)^2 + 4 b_1^2}\right)\\
        E_2 = \frac{1}{2}\left(a_0 + a_1 + \sqrt{(a_0-a_1)^2 + 4 b_1^2}\right)
    \end{aligned}
\end{equation}

Similarly \(\ket{\psi_L(s)}\) can be found by just substituting \(a_0\rightarrow a_0^*,b_1\rightarrow b_1^*,a_1\rightarrow a_1^*\).

If both \(a_0,b_1\) and \(a_1\) are real, then the complexity profiles will be periodic. It can be shown,
\begin{equation}
    \mathcal C(s) = \frac{4 b_1^2 \sin^2 ((E_1-E_2)s/2)}{(E_1-E_2)^2}
\end{equation}
early-time behavior is,
\begin{equation}
    \mathcal C(s) \approx  b_1^2 s^2
\end{equation}
and amplitude of oscillation is given by,
\begin{equation}
   \max \mathcal C(s) = \frac{4 b_1^2}{4 b_1^2 + (a_0-a_1)^2}
\end{equation}

For generic cases, when \(E_1\) and \(E_2\) are not real, then the complexity profile saturates instead of oscillating. Without loss of generality, suppose \(\Im m E_1 > \Im m E_2\), then for \(s >> 1/(\Im m E_1-\Im m E_2)\),
\begin{equation}
    \ket{\psi_R(s)} \rightarrow 
    \frac{1}{(E_2-E_1)}
    \begin{pmatrix}
        (E_2 - a_0)\\
        b_1
    \end{pmatrix}e^{-i E_1 s}
\end{equation}
and right complexity goes to,
\begin{equation}
    \mathcal C_R(s) \rightarrow \frac{|b_1|^2}{|b_1|^2+|E_2-a_0|^2}
\end{equation}
similarly, left complexity goes to,
\begin{equation}
    \mathcal C_L(s) \rightarrow \frac{|b_1|^2}{|b_1|^2+|E_1-a_0|^2}
\end{equation}
In this case early-time behaviour is also,
\begin{equation}
    \mathcal C(s) \approx |b_1|^2 s^2
\end{equation}
 
The idea of level-2 pseudo-modular spread complexity can be generalized to \textit{level-\(k\) pseudo-modular complexity} (this level-\(k\) not be confused with level-\(k\) of Chern-Simons gauge theory). In this case, one utilizes the Lanczos coefficients \(a_0,a_1,\dots,a_{k-1}\) and \(b_1,b_2,\dots,b_{k-1}\) to construct a \(k\times k\) tridiagonal matrix,
\begin{equation}
    T = 
    \begin{pmatrix}
        a_0 & b_1 & 0 & \cdots& 0 & 0 \\
        b_1 & a_1 & b_2 & \cdots& \cdot  & \cdot\\
        0 & b_2 & \cdot & \cdots& \cdot  & \cdot\\
         \cdot & \cdot & \cdot & \cdots& \cdot  & \cdot\\
        \cdot & \cdot & \cdot & \cdots& a_{k-2} & b_{k-1}\\
        \cdot & \cdot & \cdot & \cdots& b_{k-1} & a_{k-1}\\
    \end{pmatrix}
\end{equation}
To obtain these many Lanczos coefficients first \(2k-1\) moments, \(m_R^{(1)},m_R^{(2)},\dots,m_R^{(2k-1)}\), will suffice.

The solution to Eq.(\ref{eq: rec. diff eqn pseudo}) with the \(k\)-level approximation is given by,
\begin{equation}
    \ket{\psi_R (s)} =
    \begin{pmatrix}
        \phi_0^R (s)\\
        \phi_1^R (s)\\
        \cdot\\
        \cdot\\
        \phi_{k-1}^R (s)\\
    \end{pmatrix} =
    e^{-i T s} 
    \begin{pmatrix}
        1\\
        0\\
        \cdot\\
        \cdot\\
        0\\
    \end{pmatrix}
\end{equation}

This method is suitable for implementing numerically if the Lanczos coefficients or the moments are known, and can be applied for both modular and pseudo-modular spread complexities.

\section{Qubit Examples}\label{sec: qubit examples}
In this section we analyse some multi-qubit systems. Pseudo-entropy for these systems were studied before in Refs. \cite{Shinmyo_2024,parzygnat2023svdentanglemententropy}. Here we show the behavior of the capacity, pseudo-capacity and pseudo-modular spread complexity of the these multi-qubit systems.

\subsection{A. Spins with 4-body Entanglement}
Consider a 4-qubit system, and take one state to have entanglement between first two qubits and last two qubits, and the other state has 4-body entanglement between all the the qubits. Such two states are,
\begin{equation}
    \begin{aligned}
        &\ket{\psi} = (\cos\theta'\ket{00}+\sin\theta'\ket{11})\otimes (\cos\theta'\ket{00}+\sin\theta'\ket{11})\\
        &\ket{\varphi} = \cos\theta \ket{0000} + e^{i\phi}\sin\theta \ket{1111}
    \end{aligned}
\end{equation}
While calculating reduced density/transition matrices, we trace-out the 1st and 4th qubit. That is, our subsystem \(A\) consists of 2nd and 3rd qubit.
\begin{equation}\label{eq: 4-qubit reduced density matrices}
    \begin{aligned}
        &\rho_A^{\psi} = \text{Tr}_B \left(\ket{\psi}\bra{\psi}\right)\\
        &= \cos^4\theta'\ket{00}\bra{00}+\sin^4\theta'\ket{11}\bra{11}\\
&+\cos^2\theta'\sin^2\theta'(\ket{01}\bra{01}+\ket{10}\bra{10})\\
&\rho_A^{\varphi} = \text{Tr}_B \left(\ket{\varphi}\bra{\varphi}\right)\\
&= \cos^2\theta \ket{00}\bra{00} + \sin^2\theta \ket{11}\bra{11}\\
&\rho_A^{\psi|\varphi} = \text{Tr}_B \left(\ket{\psi}\bra{\varphi}\right)/\braket{\varphi|\psi}\\
&= \frac{\cos^2\theta'\cos\theta\ket{00}\bra{00}+e^{-i\phi}\sin^2\theta'\sin\theta\ket{00}\bra{00}}{\cos^2\theta'\cos\theta+e^{-i\phi}\sin^2\theta'\sin\theta}
    \end{aligned}
\end{equation}

\begin{figure}
    \centering
    \includegraphics[width=0.49\linewidth]{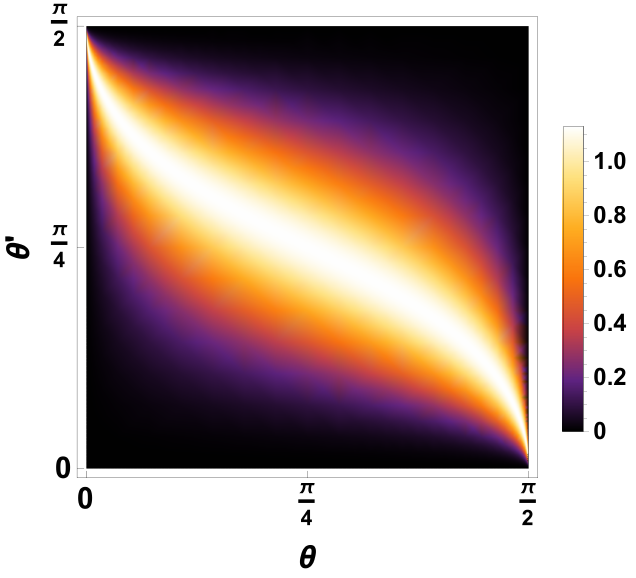}
    \includegraphics[width=0.49\linewidth]{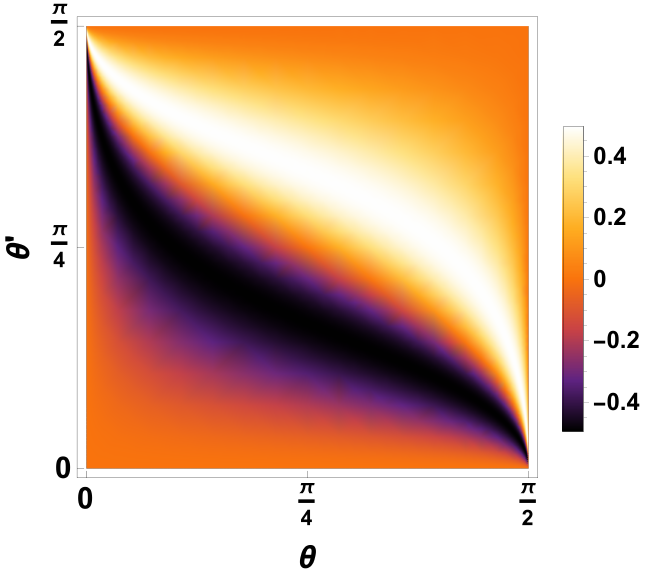}
    \includegraphics[width=0.49\linewidth]{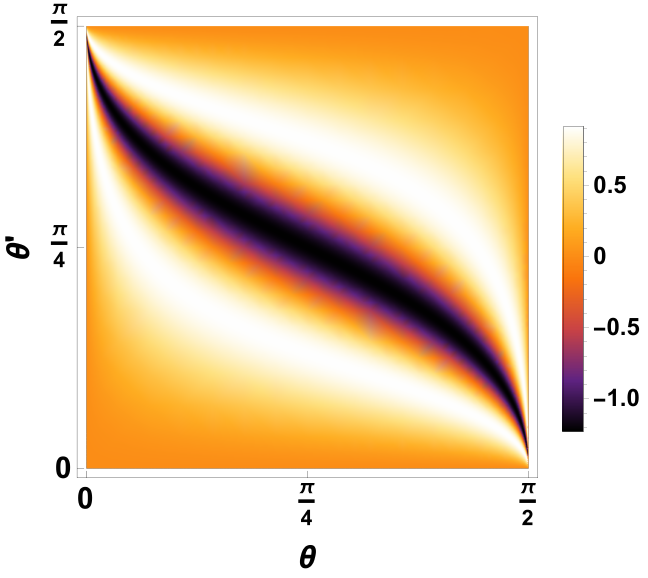}
    \includegraphics[width=0.49\linewidth]{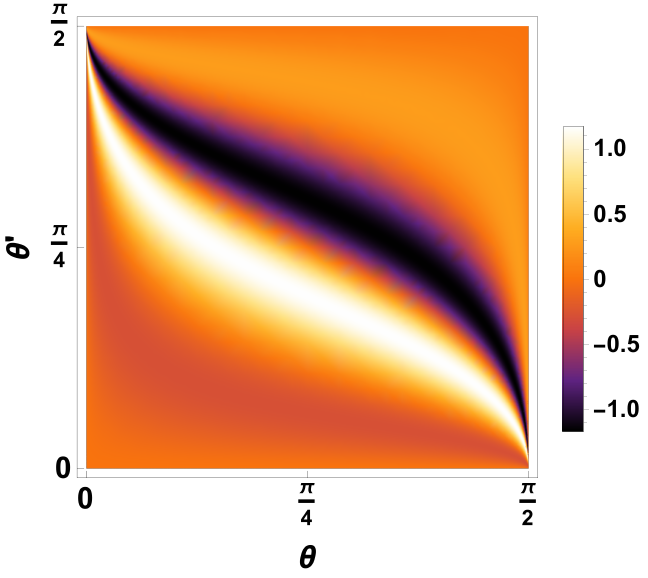}
    \caption{\textbf{Pseudo-entropy and pseudo-capacity.} We show the behavior of pseudo-entropy and pseudo-capacity for the reduced transition matrix defined in Eq.(\ref{eq: 4-qubit reduced density matrices}), as a function of \(\theta\) and \(\theta'\). The top panel shows the real (left) and imaginary (right) parts of pseudo-entropy. The bottom panel shows the real (left) and imaginary (right) parts of pseudo-capacity. In the plots, the parameters \(\phi\) is taken to be \(\pi/2\).}
    \label{fig: pseudo-entropy and capacity 4-qubit}
\end{figure}

First we calculate the pseudo-entropy and pseudo-capacity as a function of two parameters \(\theta\) and \(\theta'\). Here we fix \(\phi\) to be \(\pi/2\). Fig.~\ref{fig: pseudo-entropy and capacity 4-qubit} shows real and imaginary parts of pseudo-entropy in the top-panel and that of pseuso-capacity in the bottom-panel. From the real parts of pesudo-entropy and pseudo-capacity it is observed that, when pseudo-entropy is zero, pseudo-capacity is also zero. Pseudo-capacity is maximum in the region where there is a transition in the pseudo-entropy, which is completely analogous to what happens for Hermitian case. What is interesting is that when there is a maxima in the pseudo-entropy, pseudo-capacity is (maximum) negative, as opposed to Hermitian case where it is zero. Similar behavior is apparent in the imaginary parts of pseudo-entropy and pseudo-capacity also.

\begin{figure}
    \centering
    \includegraphics[width=0.49\linewidth]{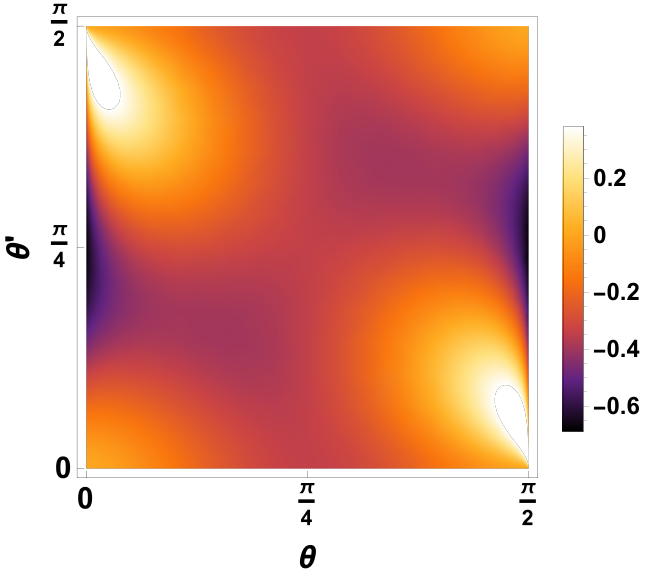}
        \includegraphics[width=0.49\linewidth]{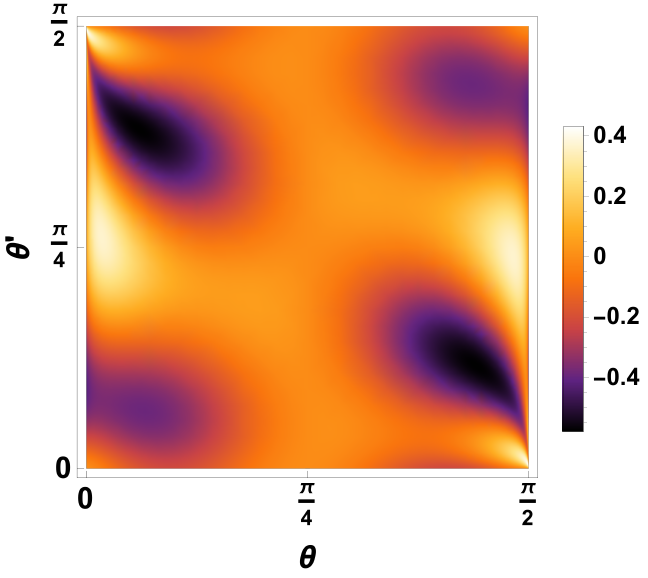}

    \caption{\textbf{Entropy excess and capacity excess.} The entropy excess \(\Delta S_E\) (defined in Eq.(\ref{eq: entropy excess})) is shown in the left plot and the capacity excess \(\Delta C_E\) (defined in Eq.(\ref{eq: capacity excess})) is shown in the right plot. For the plots, the paramater \(\phi\) is taken to \(0\). It is observed that the capacity excess has maximum value when there is a transition in the entropy excess.}
    \label{fig: entropy excess and capacity excess}
\end{figure}

Now, we focus on entropy excess \(\Delta S_E\), which is defined by,
\begin{equation}\label{eq: entropy excess}
    \Delta S_E = S_E^{\psi|\varphi} - 0.5 \times (S_E^{\psi|\psi}+S_E^{\varphi|\varphi})
\end{equation}
As observed in~\cite{Shinmyo_2024}, \(\Delta S_E > 0\) only in the regions \(\theta\) small and \(\theta'\) large or \(\theta'\) small and \(\theta\) large, see left plot of Fig.~\ref{fig: entropy excess and capacity excess}.

Similar to entropy excess, here we define capacity excess,
\begin{equation}\label{eq: capacity excess}
    \Delta C_E = C_E^{\psi|\varphi} - 0.5 \times (C_E^{\psi|\psi}+C_E^{\varphi|\varphi})
\end{equation}
As shown in right plot in Fig.~\ref{fig: entropy excess and capacity excess}, \(\Delta C_E\) has a richer structure. It is maximum when there is a transition in \(\Delta S_E\), it is (maximum) negative when \(\Delta S_E\) is maximum positive. These observations reinforce that capacity of entanglement can indeed be a important quantity for determining phase transitions.

\begin{figure}
    \centering
    \includegraphics[width=0.49\linewidth]{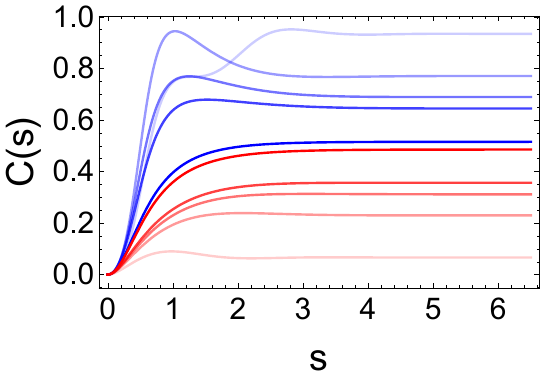}
    \includegraphics[width=0.49\linewidth]{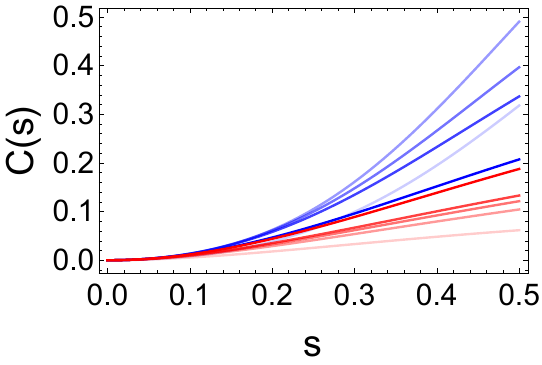}
    \caption{\textbf{Pseudo-modular spread complexity.} The blue curves denote right modular spread complexities \(\mathcal C_R^{\psi|\varphi} (s)\) and the red curves denote the left modular spread complexities \(\mathcal C_L^{\psi|\varphi} (s)\). The plots are done for \(\theta=\theta'\) line (and with \(\phi=\pi/2\)) and darker shade indicates higher value of \(\theta\), with the lighest shade for \(\theta=\pi/8\) and darkest shade for \(\theta=\pi/4-0.01.\)}
    \label{fig: pseudo-modular 4-qubit}
\end{figure}

Next we show the behavior of pseudo-modular spread complexity in Fig.~\ref{fig: pseudo-modular 4-qubit}. We have chosen \(\phi=\pi/2\), so that right (blue curves) and left (red curves) pseudo-modular spread complexity will be different. From the capacity plot (bottom-left) in Fig.~\ref{fig: pseudo-entropy and capacity 4-qubit}, we find interesting transitions along the line \(\theta=\theta'\). In Fig.~\ref{fig: pseudo-modular 4-qubit} we have plotted pseudo-modular spread complexities for some points along the line \(\theta=\theta'\). The lightest shade is used for \(\theta=\pi/8\) and the darkest shade for \(\theta=\pi/4-0.01\). The early time growth (see the left plot in Fig.~\ref{fig: pseudo-modular 4-qubit}) can be explained from the value of corresponding pseudo-capacities. One interesting observation is that there is a peak in the right modular spread complexity for the case where capacity was maximum positive. There is no peak like structure for the cases when capacity was zero (near \(\theta=\theta'\approx 0\)) or capacity was maximum negative (near \(\theta=\theta'\approx \pi/4\)). For \(\theta=\theta'=\pi/4\), the right and left modular complexities coincide and for \(\theta=\theta'>\pi/4\), \(\mathcal C_R\) and \(\mathcal C_L\) switch their roles.

\subsection{B. Ising model Phase Transition} 

Now, we shall study the paramagentic-to-ferromagnetic phase transition in 1\(d\) transverse field Ising model (TFIM), from the perspective of modular spread complexity. The Hamiltonian of this system is,
\begin{equation}
    \hat{H}(J,h) = - J \sum_{i=0}^{L-1} \sigma^z_i \sigma^z_{i+1} - h \sum_{i=0}^{L-1} \sigma^x_i
\end{equation}
here \(\sigma^z_i\) and \(\sigma^x_i\) are the Pauli matrices (at site index \(i\)) and the spins reside on a periodic lattice with period \(L\). This model has Kramers-Wannier duality~\cite{Ruelle_2005} and it can be solved exactly using Jordan-Wigner transformation. A quantum phase transition happens at the self-dual point \(J/h=1\)~\cite{Sachdev_2000}. The phase for \(h<J\) is ferromagnetic and the phase for \(h>J\) is paramagnetic. 

\begin{figure}[htbp]
    \centering
    \includegraphics[width=0.95\linewidth]{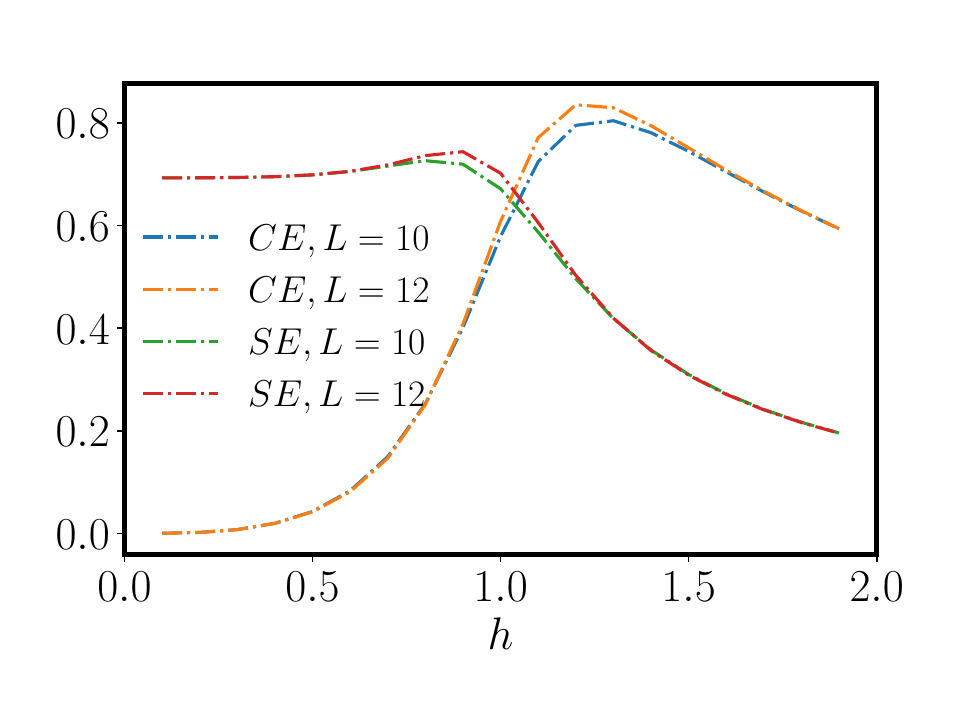}
    \caption{\textbf{Entanglement entropy and capacity of entanglement in Ising phase transition.} Half-chain entanglement entropy and capacity of entanglement in TFIM is plotted as a function of \(h\) for fixed \(J=1\).  Near the quantum phase transition point \(J=h=1\), both \(S_E\) and \(C_E\) show a transition. The plots are made for two different system sizes \(L=10,12\).}
    \label{fig: SE and CE Ising}
\end{figure}

There is transition in the ground-state entanglement property during the quantum phase transition. Half-chain entanglement entropy of the ground state is higher in the ferromagnetic phase, lower in the paramagnetic phase and it has a peak like feature during the transition, which we can observe if we plot entanglement entropy \(S_E\) as a function of \(h\) for fixed \(J=1\), see Fig.~\ref{fig: SE and CE Ising}. Here we show that there is a transition in the capacity of entanglement \(C_E\) also (which determines early time growth of modular spread complexity). It is observed that \(C_E\) has a peak near the transition, and it decreases in both the phases. Interestingly, the curves of \(S_E\) and \(C_E\) intersects at almost \(h=1\). This shows that \(C_E\) is also an important measure for determining quantum phase transtions.

\begin{figure}
    \centering
    \includegraphics[width=0.95\linewidth]{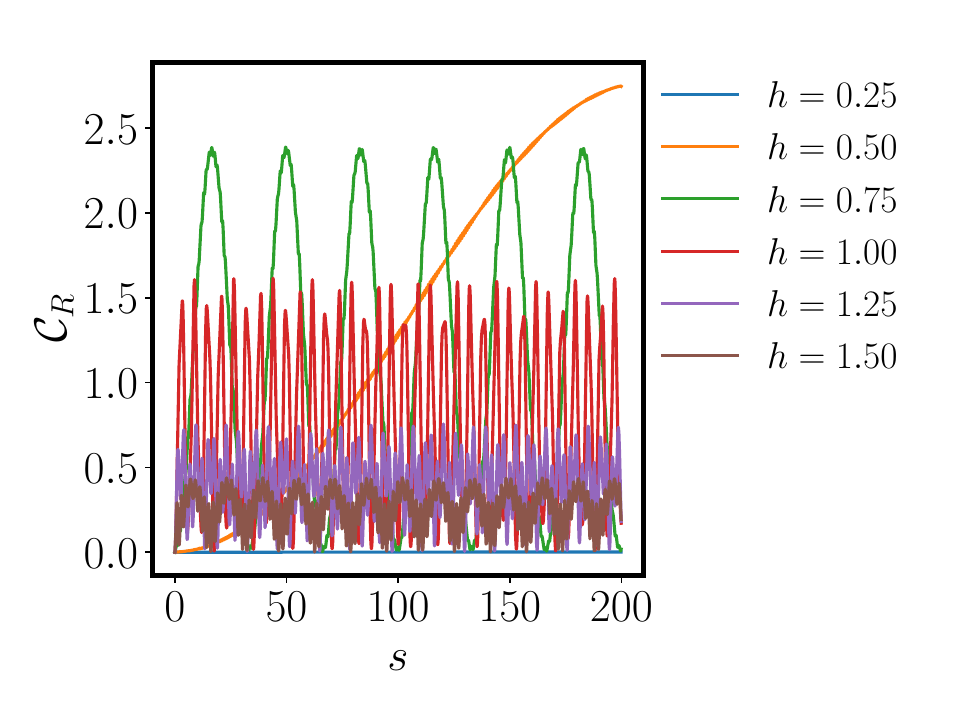}
    \caption{\textbf{Modular spread complexity in Ising phase transition.} We show the behavior of modular spread complexity for the ground state density matrix with half of the chain traced out. The value of \(J\) is fixed to 1 and the plots are made for different values of \(h\). It is observed that, the modular spread complexity has distinctive profiles in different phases, and in particular, near the quantum phase transition, a highly oscillating profile.}
    \label{fig: modular spread complexity Ising}
\end{figure}

while \(C_E\) just gives early time measure of modular spread complexity, there is a significant difference between the two phases from the perspective of full modular spread complexity profile. Fig.~\ref{fig: modular spread complexity Ising} shows the profiles of (right) modular spread complexity for different values of \(h\) at fixed \(J=1\). In all cases the modular spread complexity has an oscillating profile. In the ferromagnetic phase the oscillation frequency is less and amplitude is more. In the paramagnetic phase, the oscillation amplitude is smaller. Finally, near the quantum phase transition, oscillation frequency becomes very high.

\begin{figure}
    \centering
    \includegraphics[width=0.49\linewidth]{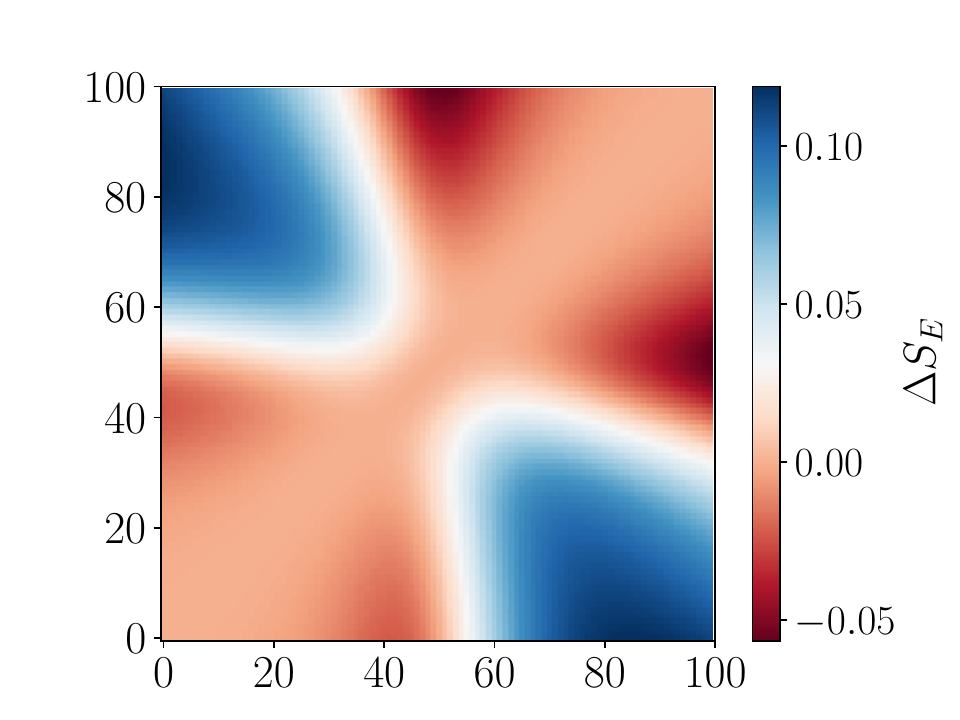}
    \includegraphics[width=0.49\linewidth]{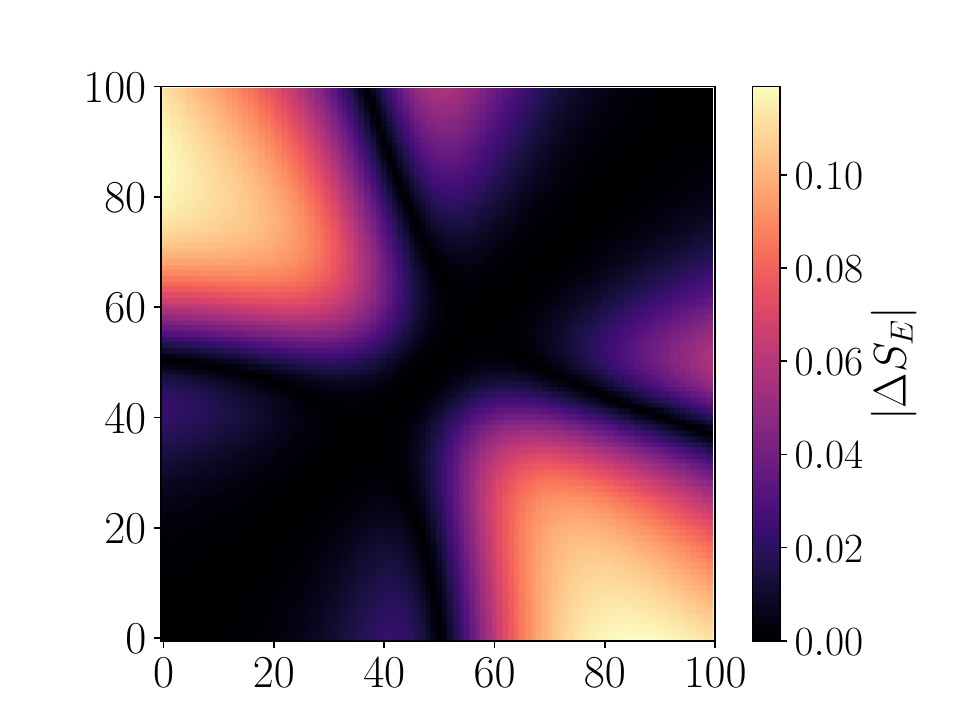}

    \includegraphics[width=0.49\linewidth]{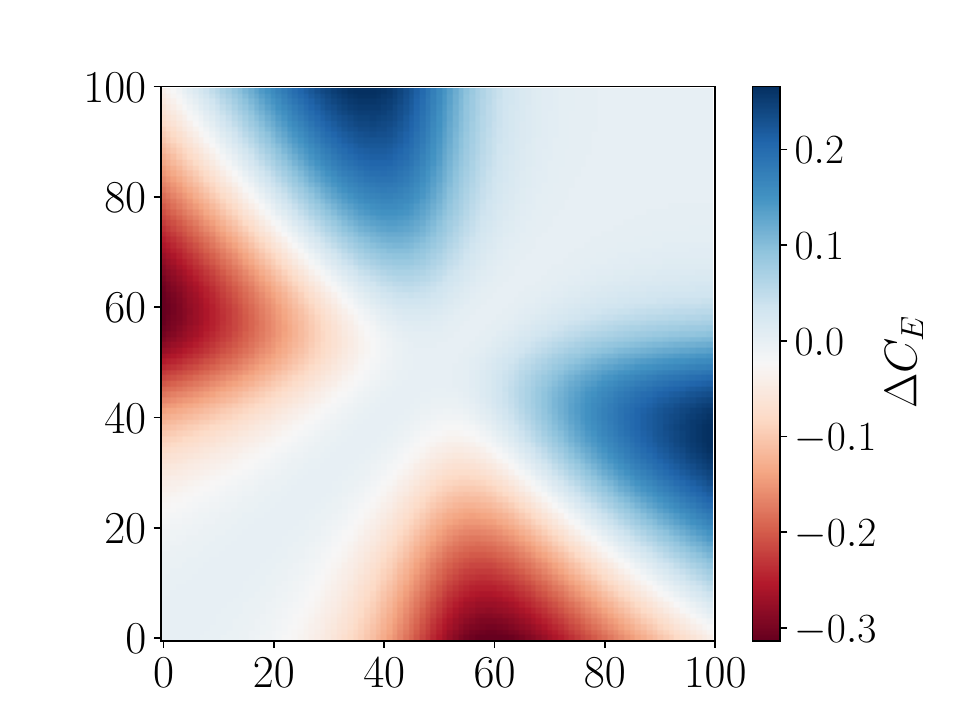}
    \includegraphics[width=0.49\linewidth]{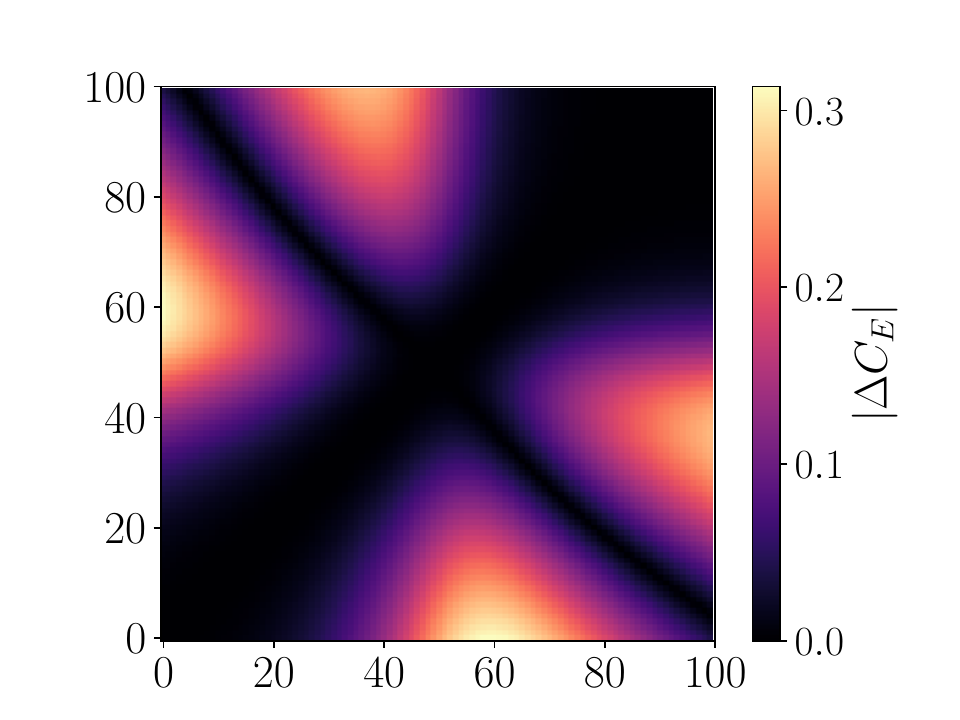}
    
    \caption{\textbf{Entropy excess and capacity excess in TFIM} The top panel shows the behavior of \(\Delta S_E\) in the \(h_1-h_2\) plane and the bottom panel shows the behavior of \(\Delta C_E\) in the \(h_1-h_2\) plane. In the above plots, the axes are graduated according to an index running from 0 to 100. The value of \(h\) corresponding to an index \(n\) is given by, \(h = 0.25 + 0.015\,n\)}
    \label{fig: entropy and capacity excess TFIM}
\end{figure}

Now we turn to pseudo-entropy, pseud-capacity and pseudo-modular complexity for TFIM. The setup is following. We consider the TFIM Hamiltonian for two parameter choices, \((J_1,h_1)\) and \((J_2,h_2)\). Call the corresponding ground states \(\ket{\Omega_{J_1,h_1}}\) and \(\ket{\Omega_{J_2,h_2}}\) respectively. Depending on the parameter choices, both the ground states can either be in same phase or in different phases. The transition matrix \(\rho^{\psi|\varphi}\) is constructed by choosing,
\begin{equation}\label{eq: ground states TFIM}
    \ket{\psi} = \ket{\Omega_{J_1,h_1}},\,\,\,\ket{\varphi} = \ket{\Omega_{J_2,h_2}}
\end{equation}
The susbsyetem \(A\) is again chosen to be the half of the chain.
It was observed earlier that always \(\Delta S_E \leq 0\) if the two ground states are in a same phase. If the ground states are in a different phase, then the inequality may be violated. The top panel in Fig.~\ref{fig: entropy and capacity excess TFIM}  depicts this observation.

The key point to note from the bottom panel of Fig.~\ref{fig: entropy and capacity excess TFIM} is that the capacity excess has higher magnitude whenever there is a transition in the entropy excess. In particular, when one of the ground states is near the quantum phase transition and the other is not, the capacity excess has higher magnitude. Recall that, capacity \(C_E\) also has a maxima near the quantum phase transition.
On the other hand, when both of the ground states are near quantum phase transition, capacity excess is zero, so is the entropy excess. 

\begin{figure}
    \centering
    \includegraphics[width=0.95\linewidth]{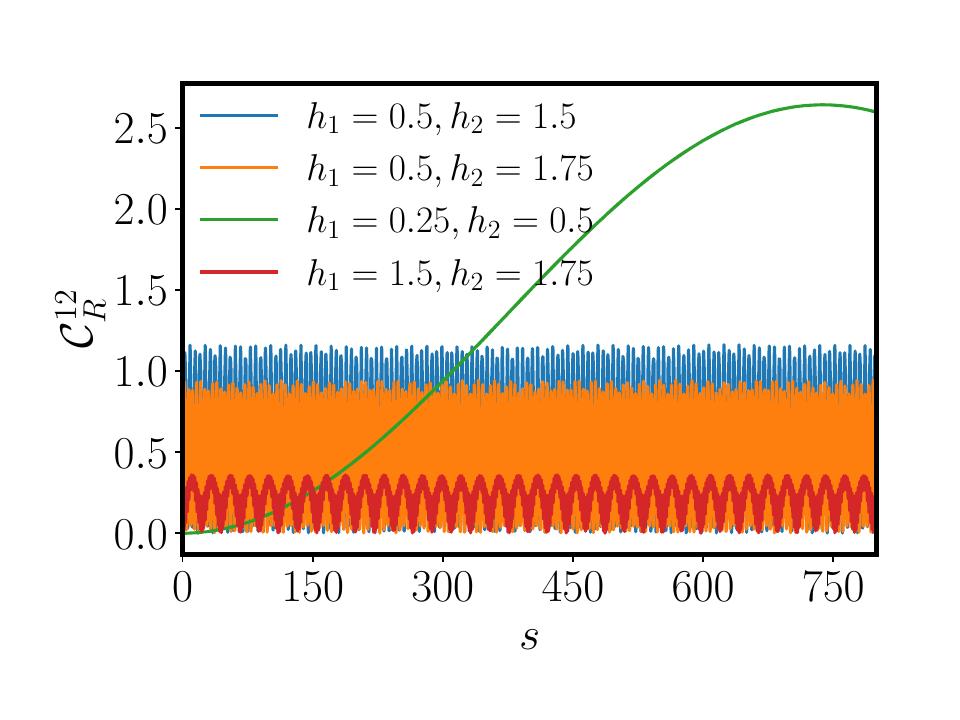}
    \caption{\textbf{Pseudo-modular spread complexity in TFIM} The (right) pseudo-modular spread complexity is plotted for different choices of \(h_1\) and \(h_2\) with fixed \(J_1=J_2=1\). It is observed that the complexity profile is highly oscillating when the two ground states in Eq.(\ref{eq: ground states TFIM}) belong to different phases.}
    \label{fig: pseudo-modular spread complexity Ising}
\end{figure}

Now we calculate pseudo-modular complexity corresponding to the non-Hermitian density matrix \(\rho_A^{\psi|\varphi}\) where \(\ket{\psi}\) and \(\ket{\varphi}\) are defined in Eq.~(\ref{eq: ground states TFIM}). We call the pseudo-modular spread complexity coming from \(\rho_A^{\psi|\varphi}\) as \(\mathcal C^{12}\). We plot the right pseudo-modular spread complexity \(\mathcal C^{12}_R\) in Fig.~\ref{fig: pseudo-modular spread complexity Ising} for different choices of \(h_1\) and \(h_2\) (\(J_1\) and \(J_2\) are fixed to 1). From plots we see that the pseudo-modular complexity has qualitatively similar profile as the corresponding modular spread complexities when \(h_1\) and \(h_2\) correspond to the same phase. However, when \(h_1\) and \(h_2\) correspond to different phases, the pseudo-modular complexity profile is similar to modular spread complexity profile for the system near quantum phase transition.

\section{Chern-Simons gauge theory}\label{sec: Chern-Simons gauge theory}

Three-dimensional Chern-Simons gauge theory~\cite{Labastida_1999,dunne1999aspectschernsimonstheory,Zanelli:2010zz,Witten:1992fb} is an example of a topological field theory~\cite{BIRMINGHAM1991129} (that is, the action does not depend on the metric of the underlying manifold). Entanglement entropy as well as pseudo-entropy in this topological field theory can be calculated by using the surgery method~\cite{Witten:1988hf,Dong:2008ft}. Our goal is to understand the behavior of modular and pseudo-modular spread complexity in Chern-Simons gauge theory with Wilson lines. The (Euclidean) action of Chern-Simons gauge theory on a 3d manifold \(\mathcal M\), with gauge group \(SU(N)\) and (integer) level \(k\) is,
\begin{equation}
    S_{CS}[A] = - i \frac{k}{4\pi} \int_{\mathcal M} \text{Tr}\left( A \wedge d A + \frac{2}{3} A \wedge A \wedge A\right)
\end{equation}
where \(A\) is connection one-form.
Wilson loops are gauge-invariant operators in the theory, and to specify a Wilson loop around a closd path \(\mathcal C\) we have to specify a particular representation \(R\) (of \(SU(N)\)),
\begin{equation}
    W_R[A] = \text{Tr}_R \,\mathcal P \exp\left( \int_{\mathcal C} A\right)
\end{equation}
where \(\mathcal P\) denotes path ordering.

Partition function in Chern-Simons gauge theory with Wilson loops can be evaluated by using its duality with \(\widehat{SU(N)_k}\) Wess-Zumino-Witten (WZW) model, which is a 2\(d\) CFT. The results of partition function involves the modular \(S\)-matrix elements, which is defined as follows. If we put the WZW model on a Torus, then by the modular invariance of the theory, the character \(\chi_i (\tau)\) (where \(\tau\) is the complex structure on the torus) will have the following transformation under \(\tau \rightarrow -1/\tau\),
\begin{equation}
    \chi_i (-1/\tau) = \sum_j \mathcal S^j_i\, \chi_j (\tau)
\end{equation}
here \(\mathcal S_i^j\) are the elements of modular \(S\)-matrix \(\mathcal S\), which is symmetric and unitary. Quantum dimension \(d_i\) of a representation \(i\) is defined by \(\mathcal S_i^0/\mathcal S_0^0\), with total quantum dimension \(\mathcal D\) being, \(\mathcal D = \sqrt{\sum_i |d_i|^2} = 1/\mathcal S_0^0\). For example, the modular \(S\)-matrix for \(\widehat{SU(2)_k}\) WZW model is,
\begin{equation}
    \mathcal S_{i}^j = \sqrt{\frac{2}{k+2}} \sin \left[\frac{\pi (2i+1)(2j+1)}{k+2}\right]
\end{equation}
The partition function on a 3-sphere \(\mathbb S^3\) without any Wilson loops is, \(Z[\mathbb S^3]= \mathcal S_0^0 = 1/\mathcal D\); partition function on \(\mathbb S^3\) with a single Wilson loop \(W_{R_i}\) is, \(Z[\mathbb S^3,R_i]= \mathcal S_0^i\); partition function on \(\mathbb S^3\) with two disconnected Wilson loops \(W_{R_i}\) and \(W_{R_j}\) is, \(Z[\mathbb S^3,R_i,R_j]= \mathcal S_0^i \mathcal S_0^j/\mathcal S_0^0\) and partition function on \(\mathbb S^3\) with two linked Wilson loops \(W_{R_i}\) and \(W_{R_j}\) is, \(Z[\mathbb S^3,L(R_i,R_j)]= \mathcal S_i^j \). The results for different partition functions quoted in this section can be found in~\cite{Nishioka_2021,Dong:2008ft}.

\subsection{A. Topological pseudo-modular complexity on \(\mathbb S^2\)}
We consider Chern-Simons gauge theory on the manifold \(\mathbb B^3\), whose boundary is \(\mathbb S^2\), on which some (anyonic) excitations will be created. We take subsystem \(A\) to be half of \(\mathbb S^2\) and subsystem \(B\) to be remaining half. First, we will give examples for which topological capacity of entanglement as well as topological modular complexity vanish identically.

Suppose we create two excitations, one belonging to representation \(R_i\) and another belonging to \(\Bar{R_i}\) of \(\widehat{SU(N)_k}\), where the first excitation is in \(A\) and the second is in \(B\). It can be shown that \(\text{Tr}_A \rho_A^n\) can be obtained from partition functions on \(\mathbb S^3\) with Wilson loop. In particular,
\begin{equation}
    \text{Tr}_A \rho_A^n = \frac{Z[\mathbb S^3, R_i]}{Z[\mathbb S^3, R_i]^n}
\end{equation}
By direct calculation, we obtain the moments as,
\begin{equation}
    m^{(k)} = \log^k (Z[\mathbb S^3, R_i])
\end{equation}
which implies that only \(a_0\) (which is the von-Neumann entropy) is non-zero and all the other Lanczos coefficients (including \(b_1^2\), which is the capacity of entanglement) are zero. Therefore, \textit{topological} modular complexity will vanish identically. 

Next, we create four excitations on \(\mathbb S^2\), two of them in fundamental representation and two of them in anti-fundamental representation. In the first case, \(A\) contains one fundamental \(j\), one anti-fundamental \(\Bar{j}\) and similar of \(B\). Still, there are two possible configuration of Wilson lines, (a) \(j,\Bar{j}\) in \(A\) are connected by a Wilson line and similar for \(B\), which we call the state \(\ket{\psi}\), (b) \(j\) of \(A\) is connected with \(\Bar{j}\) of \(B\) and vice-versa, which we call the state \(\ket{\varphi}\). For these, the results for reduced density/transition matrix are,
\begin{equation}
    \begin{aligned}
        \text{Tr}_A(\rho_A^{\psi})^n &= (\mathcal S_0^0)^{1-n}\\
        \text{Tr}_A(\rho_A^{\varphi})^n &= \left[\frac{(\mathcal S_0^j)^2}{\mathcal S_0^0}\right]^{1-n}\\
        \text{Tr}_A(\rho_A^{\psi|\varphi})^n &= (\mathcal S_0^0)^{1-n}
    \end{aligned}
\end{equation}
Again all the Lanczos coefficients except \(a_0\) will vanish, and we will get vanishing modular and pseudo-modular spread complexity for these cases too.

In the second case, sub-region \(A\) will contain two \(j\)'s and sub-region \(B\) will contain two \(\Bar{j}\)'s. Therefore, there will be two Wilson lines connecting the sub-regions \(A\) and \(B\). There are countably infinite number of possibilities based on how many times these two Wilson lines twist around each other. The states corresponding to these configurations are called \(\ket{\psi_a}\), where the integer \(a\) denotes the number of twists, and negative \(a\) implies twist in the opposite sense. We consider reduced transition matrix \(\rho_A^{a|b} = \text{Tr}_B(\ket{\psi_a}\bra{\psi_b})/\braket{\psi_b|\psi_a}\), and the corresponding result is,
\begin{equation}
    \begin{aligned}
        &\text{Tr}_A \left[\left(\rho_A^{a|b}\right)^n\right]\\
        &= (\mathcal S_0^0[N])^{1-n} \frac{(q^{\frac{1}{2}})^{|a-b|n}\frac{[N+1]}{[2]}+(-q^{\frac{1}{2}})^{|a-b|n}\frac{[N-1]}{[2]}}{\left[(q^{\frac{1}{2}})^{|a-b|}\frac{[N+1]}{[2]}+(-q^{\frac{1}{2}})^{|a-b|}\frac{[N-1]}{[2]}\right]^n}\\
        &= (\mathcal S_0^0[N])^{1-n} \frac{\alpha^n \Tilde{\alpha}+\beta^n \Tilde{\beta}}{(\alpha \Tilde{\alpha}+\beta \Tilde{\beta})^n}
    \end{aligned}
\end{equation}
where \(q=e^{\frac{2\pi i}{N+k}}\) and,
\begin{equation}
\begin{aligned}
    &[x] = \frac{q^{x/2}-q^{-x/2}}{q^{1/2}-q^{-1/2}}\\
    &\alpha = (q^{1/2})^{|a-b|},\,\,\,\beta = (-q^{-1/2})^{|a-b|}\\
    &\Tilde{\alpha} = \frac{[N+1]}{[2]},\,\,\,\Tilde{\beta} = \frac{[N-1]}{[2]}
\end{aligned}
\end{equation}
The case \(a=b\) reduces to,
\begin{equation}
    \text{Tr}_A \left[\left(\rho_A^{a|a}\right)^n\right] = \left(\mathcal S_0^0 [N]^2\right)^{1-n}
\end{equation}
So, the case \(a=b\) will give vanishing modular spread complexity always. But when \(a\neq b\) we will get non-trivial pseudo-modular complexity, which we calculate now.

For general \(a\neq b\) case, the moments can be written as,
\begin{equation}\label{eq: moments Wilson on S2, T2}
    m^{(k)} = (-1)^k \sum_{r=0}^k {}^k C_r \braket{r} (-x)^{k-r}
\end{equation}

where,
\begin{equation}
    x = \log \left(\mathcal S_0^0 [N] (\alpha \Tilde{\alpha}+\beta \Tilde{\beta})\right),\,\,\braket{n} = \frac{\log^n( \alpha) \alpha \Tilde{\alpha}+\log^n (\beta) \beta \Tilde{\beta}}{\alpha \Tilde{\alpha}+\beta \Tilde{\beta}}
\end{equation}

By direct calculation from the moments we get the first few Lanczos coefficients (all the Lanczos coefficients can be determined systematically using moment recursion algorithm~\cite{Nandy:2024htc}) as,
\begin{equation}\label{eq: pseudo-lanczos S2}
    \begin{aligned}
        &a_0 = x - \braket{1}\\
        &b_1 = \sqrt{\braket{2}-\braket{1}^2}\\
        &a_1 = x - \frac{\braket{1}^3+\braket{3}-2\braket{1}\braket{2}}{\braket{2}-\braket{1}^2}
    \end{aligned}
\end{equation}

using these expressions, we can calculate the level-2 pseudo-modular spread complexity introduced in section [ref] for \(\widehat{SU(2)_k}\) WZW model, for which \(\mathcal S_0^0 = \sqrt{\frac{2}{2+k}}\sin\left[\frac{\pi}{k+2}\right]\).

\begin{figure}
    \centering
    \includegraphics[width=0.95\linewidth]{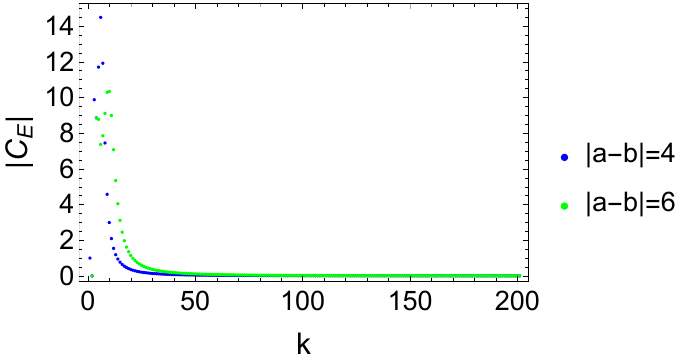}
    \includegraphics[width=0.95\linewidth]{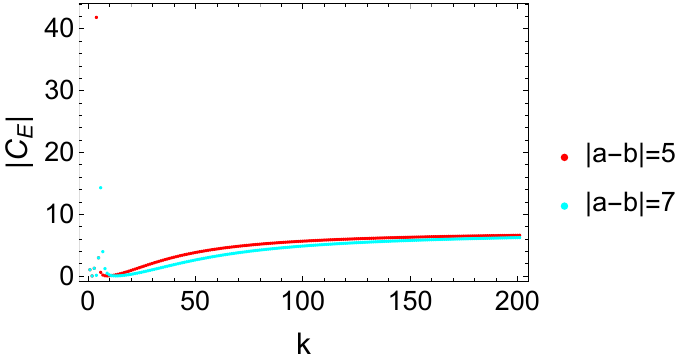}
    \caption{\textbf{Pseudo-capacity for different Wilson line configurations in \(\mathbb B^3\).} The top plot shows the modulus of pseudo-capacity for the case \(|a-b|\)=even and the bottom plot shows the case \(|a-b|\)=odd. Non-trivial capacity is observed only for the case \(|a-b|\)=odd at large \(k\). Here we have considered \(SU(2)\) Chern-Simons gauge theory.}
    \label{fig: pseudo capacity on S2}
\end{figure}

From Fig.~\ref{fig: pseudo capacity on S2}, one can observe that for \(|a-b|=\)even, the pseudo-capacity goes to zero for large \(k\), therefore the level-2 complexity profile also goes to zero. On the other hand for \(|a-b|=\)odd, the pseudo-capacity approaches a constant value depending on \(N\), which implies that the complexity approaches a fixed non-trivial profile. To explain this, note that when \(|a-b|\)=even, the pair of excitations connected are same in \(\ket{\psi_a}\) and \(\ket{\psi_b}\). Therefore, in the classical limit \(k\rightarrow \infty\), the states are the same. On the other hand, the states individually has vanishing capacity. Therefore, the pseudo-capacity in the classical limit for the \(|a-b|\)=even case should go to zero, exactly what has been observed. In contrast, for the \(|a-b|\)=odd case, the pairs of excitations connected are not same for the two states, and that there is an additional entanglement swapping~\cite{Nishioka_2021}. Therefore, the non-zero value of pseudo-capacity (consequently pseudo-modular spread complexity) can be understood as a signature entanglement swapping.

\subsection{B. Topological pseudo-modular complexity on \(\mathbb T^2\) with Wilson loops}

In this subsection we consider solid torus \(\mathbb T^2\) with a Wilson loop with representation \(R_i\). Path integral on this manifold creates the state \(\ket{R_i}\).
Let's take the subsystem \(A\) to be to a cylinder on the surface. Now, lets \(\ket{\psi}\) and \(\ket{\varphi}\) be two states with different superpositions of Wilson loops,
\begin{equation}
\begin{aligned}
    &\ket{\psi} = \sum_i \psi_i \ket{R_i}\\
    &\ket{\varphi} = \sum_i \varphi_i \ket{R_i}
\end{aligned}
\end{equation}
From Ref.~\cite{Nishioka_2021},
\begin{equation}
\text{Tr}_A\left[\left(\rho_A^{\psi|\varphi}\right)^n\right] = \frac{\sum_i (\varphi_i^*\psi_i)^n (\mathcal S_0^i)^{2(1-n)}}{(\sum_i \varphi_i^*\psi_i)^n}
\end{equation}
From here it follows that the Lanczos coefficients take same form as in Eq.~(\ref{eq: pseudo-lanczos S2}), with 
\begin{equation}\label{eq: x and <n> for T2}
    \begin{aligned}
       x &= \log \left(\sum_i \varphi_i^* \psi_i\right)\\
       \braket{n} &= \frac{\sum_i \varphi_i^* \psi_i \log^n \left(\varphi_i^* \psi_i/ (\mathcal S_0^i)^2\right)}{\sum_i \varphi_i^* \psi_i}
    \end{aligned}
\end{equation}

We illustrate this case using the \(\widehat{SU(2)_k}\) WZW model with level \(k=2\), which correspond to \(SU(2)\) Chern-Simons gauge theory at level \(k=2\). In this case two representations of Wilson loops are possible, namely \(\ket{R_0}\) and \(\ket{R_1}\). We can take the states \(\ket{\psi}\) and \(\ket{\varphi}\) as,
\begin{equation}
    \begin{aligned}
        \ket{\psi} = \sqrt{p_1}\ket{R_0} + \sqrt{1-p_1}\ket{R_1}\\
        \ket{\varphi} = \sqrt{p_2}\ket{R_0}+\sqrt{1-p_2}\ket{R_1}
    \end{aligned}
\end{equation}

\begin{figure}
    \centering
    \includegraphics[width=0.49\linewidth]{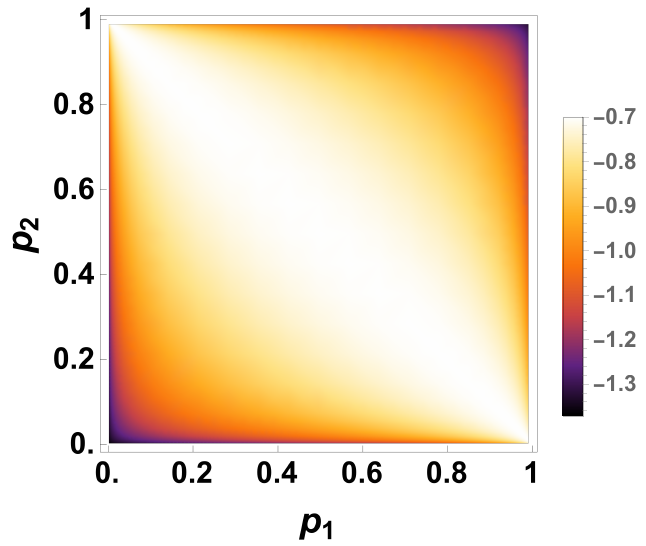}
    \includegraphics[width=0.49\linewidth]{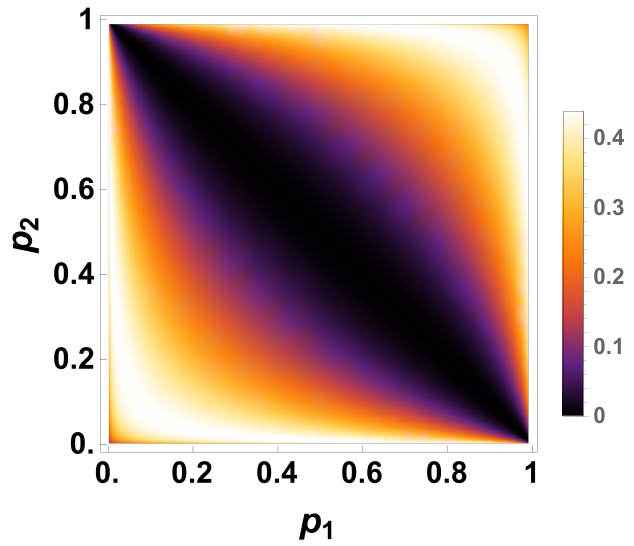}
    \caption{\textbf{Pseudo-entropy and pseudo-capacity for Wilson loop configurations on solid torus.} \textbf{Left:} pseudo-entropy, \textbf{Right:} pseudo-capacity, for \(SU(2)\) Chern-Simons gauge theory at level \(k=2\) on a solid torus with superpositon of Wilson loops belonging to two different representations.}
    \label{fig: pseudo-entropy and capacity on T2}
\end{figure}

Fig.~\ref{fig: pseudo-entropy and capacity on T2} shows the density plots of pseudo-entropy and pseudo-capacity in the \(p_1-p_2\) plane. It is observed that the pseudo-capacity is minimum in the region where the pseudo-entropy is maximum and pseudo-capacity is maximum in the region when there is a transition in the pseudo-entropy. It is consistent with observations made for the other examples analysed before. Another observation is that, the pseudo-capacity is zero along the line \(p_1+p_2 =1\).

\begin{figure}
    \centering
    \includegraphics[width=0.49\linewidth]{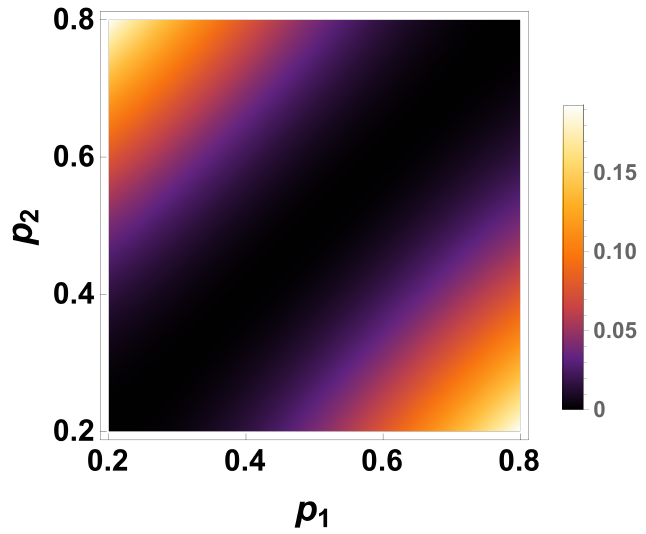}
    \includegraphics[width=0.49\linewidth]{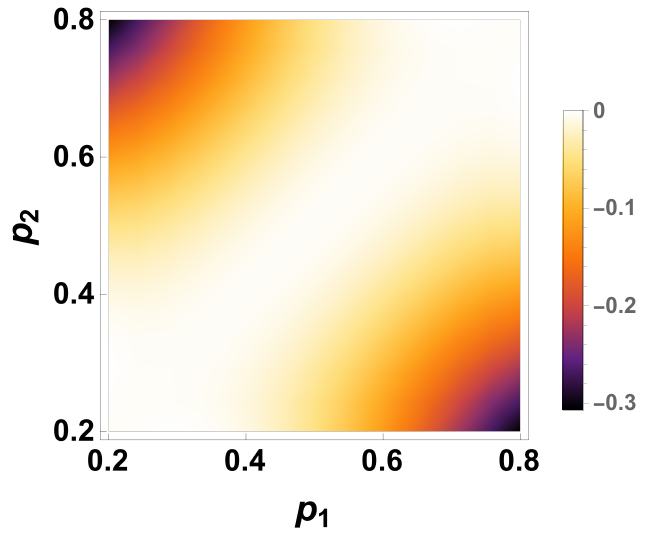}
    \caption{\textbf{Entropy and capacity excess for Wilson loop configurations on solid torus.}\textbf{Left:} entropy excess, \textbf{Right:} capacity excess, for \(SU(2)\) Chern-Simons gauge theory at level \(k=2\) on a solid torus with superpositon of Wilson loops belonging to two different representations.}
    \label{fig: entropy and capacity excess on T2}
\end{figure}

Next, we try to understand the behaviors of entropy excess \(\Delta S_E\) and capacity excess \(\Delta C_E\). Fig.~\ref{fig: entropy and capacity excess on T2} shows the density plots of entropy excess and capacity excess. It can be seen that \(\Delta S_E\) has higher positive value when the states \(\ket{\psi}\) and \(\ket{\varphi}\) are more different from each other (that is, asymmetry in \(p_1,p_2\)). On the other hand, \(\Delta C_E\) is negative and has higher magnitude in those regions.

\begin{figure}
    \centering
    \includegraphics[width=0.95\linewidth]{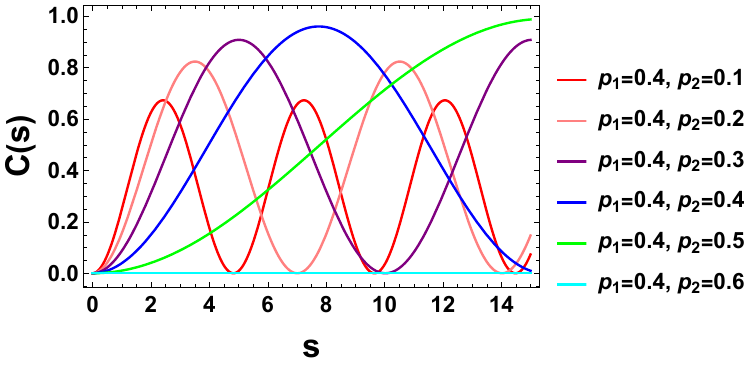}
    \caption{\textbf{Level-2 pseudo-modular spread complexity for Wilson loop configurations on solid torus.} We consider the case \(p_1=0.4\) and different values of \(p_2\). For the example considered, the Lanczos coefficients are real, therefore, the right and left pseudo-modular spread complexities coincide. The early time growth is dictated by the pseudo-capacity, but the higher Lanczos coefficients contribute to the full profile.}
    \label{fig: pseudo-modular T2}
\end{figure}

Pseudo-modular spread complexity profile can be obtained by using the moments from the expressions Eq.(\ref{eq: moments Wilson on S2, T2}) and Eq.(\ref{eq: x and <n> for T2}), computing the Lanczos coefficients and applying the method described in Sec.~\ref{sec: psedu-modular}. Fig.~\ref{fig: pseudo-modular T2} illustrates the case of level-2 psuedo-modular spread complexity which utilizes \(a_0,b_1\) and \(a_1\). Higher level pseudo-modular spread complexities will have richer structure.

\section{Discussion and Future Directions}\label{sec: discussion}

In this paper, we have extended the notion of modular spread complexity to incorporate the case of non-Hermitian density matrices which appear naturally for the case of reduced transition matrices. We have introduced the concept of right and left modular spread complexities for these generalized scenarios and described the algorithms to compute these quantities. Applying to 2-level systems, we have found that pseudo-modular spread complexity can have a saturation value, in sharp contrast to modular complexity (for Hermitian cases) which is always oscillating for 2-level systems. We have analysed the case of a transition matrix with one of the states having 2-body entanglement and another state having 4-body entanglement. In the case of TFIM, the transition matrix is constructed out of two different ground states pertaining to different model parameters. The general lesson is that higher positive value of pseudo-capacity is indicative of a transition in the pseudo-entropy. From the pseudo-modular spread complexity perspective, there is a highly oscillating behavior of complexity when the two ground states belong to different quantum phases. Therefore, pseudo-modular spread complexity can be an important tool for characterizing quantum phase transitions. Obviously, it has to be investigated in various other systems for a better understanding of its usefulness. 

Applying these concepts in Chern-Simons gauge theory with Wilson loop configurations, it is found that there exists examples where modular complexity vanishes identically but the states can have non-trivial pseudo-capacity (and hence non-trivial pseudo-modular spread complexity), which is topological in nature, in the sense that it depends on the assymmetry in the parity of the number of crossings among the Wilson lines present in the two states used for the construction of the transition matrix.

The methods introduced here can be applied in variety of other contexts, where the analysis of pseudo-modular spread complexity can be useful to understand the richer physics encoded in the (pseudo) entanglement spectrum. Some of such interesting contexts are, \(2d\) CFTs, CFTs under local quench~\cite{Shinmyo_2024}, holographic pseudo-entropy~\cite{Nishioka_2021,Kanda_2024}, time-like entanglement entropy~\cite{Doi_2023} etc. We leave these for future investigations and we hope our method can potentially unravel some interesting aspects of these systems.

\section*{Acknowledgements}

M.G. is supported by the Integrated PhD fellowship of the Indian Institute of Science. A.J. is supported by the INSPIRE fellowship by DST.

\bibliography{ref}

\end{document}